\documentclass[11pt]{article}

\usepackage{latexsym,amsmath,amssymb,theorem,epsfig,multirow,bbm}
\usepackage[noconfig]{refstyle}
\usepackage[dvipsnames]{xcolor}
\usepackage[hyperfootnotes=false, linktocpage=true, colorlinks, citecolor=blue, linkcolor=blue, urlcolor=Maroon]{hyperref}

\newlength{\xtrawidth}
\newlength{\xtraheight}
\DeclareFontFamily{OT1}{rsfs10}{}
\DeclareFontShape{OT1}{rsfs10}{m}{n}{ <-> rsfs10 }{}
\DeclareMathAlphabet{\mathscript}{OT1}{rsfs10}{m}{n}

\numberwithin{equation}{section}

\newcommand{\pt}{\partial}

\def\fnote#1#2{\begingroup\def\thefootnote{#1}\footnote{#2}
     \addtocounter{footnote}{-1}\endgroup}

\def\a{\alpha}
\def\b{\beta}
\def\g{\gamma}

\def\d{\delta}
\def\e{\epsilon}

\def\k{\kappa}
\def\l{\lambda}
\def\m{\mu}

\def\o{\omega}

\def\r{\rho}

\def\x{\xi}

\def\G{\Gamma}

\def\cA{{\mathcal A}}

\def\cL{{\mathcal L}}
\def\cK{{\mathcal K}}

\def\cN{{\mathcal N}}
\def\cO{{\mathcal O}}

\def\gsim{ \lower .75ex \hbox{$\sim$} \llap{\raise .27ex \hbox{$>$}} }
\def\lsim{ \lower .75ex \hbox{$\sim$} \llap{\raise .27ex \hbox{$<$}} }
\def\be{\begin{equation}}
\def\ee{\end{equation}}
\def\bea{\begin{eqnarray}}
\def\eea{\end{eqnarray}}

\def \a {\alpha}

\def \a {\alpha }
\def \b {\beta }
\def \o{\omega }
\def \eps {\epsilon}

\def \l {\lambda}

\def \ed {\end{document}}

\begin{document}

\begin{titlepage}

\title{{\LARGE\bf   Holomorphic Yukawa Couplings for Complete Intersection Calabi-Yau Manifolds}\\[1em] }
\author{
Stefan Blesneag${}^{1}$,
Evgeny I. Buchbinder${}^{2}$,
Andre Lukas${}^{1}$
}

\date{}
\maketitle
\begin{center} { 
${}^1${\it Rudolf Peierls Centre for Theoretical Physics, Oxford University,\\
1 Keble Road, Oxford, OX1 3NP, U.K.\\[3mm]
${}^2$ The University of Western Australia, \\
35 Stirling Highway, Crawley WA 6009, Australia
}}\\
\end{center}

\fnote{}{stefan.blesneag@wadh.ox.ac.uk}
\fnote{}{evgeny.buchbinder@uwa.edu.au}
\fnote{}{lukas@physics.ox.ac.uk} 

\vskip 1cm

\begin{abstract}
\noindent We develop methods to compute holomorphic Yukawa couplings for heterotic compactifications on complete intersection Calabi-Yau manifolds, generalising results 
of an earlier paper for Calabi-Yau hypersurfaces. Our methods are based on constructing the required bundle-valued forms explicitly and evaluating the relevant 
integrals over the projective ambient space. We also show how our approach relates to an earlier, algebraic one to calculate the holomorphic Yukawa couplings. 
A vanishing theorem, which we prove, implies that certain Yukawa couplings allowed by low-energy symmetries are zero due to topological reasons. 
To illustrate our methods, we calculate Yukawa couplings for $SU(5)$-based standard models on a co-dimension two complete intersection manifold.
\end{abstract}

\thispagestyle{empty}

\end{titlepage}

\tableofcontents
\newpage

\section{Introduction}
Advances in heterotic model building on Calabi-Yau manifolds over the past few years~\cite{Candelas:1985en}--\cite{Buchbinder:2014qda} have led to a large number of models with a realistic standard-model spectrum. It is now possible to construct models with phenomenologically promising spectra systematically searching through classes of compactifications. An immediate next step on the path towards a fully realistic particle physics model from string theory is the calculation of Yukawa couplings. 

Unfortunately, calculating Yukawa couplings for geometric compactifications of the heterotic string with general 
vector bundles is not straightforward, even at the perturbative level, and relatively few techniques and results are 
known~\cite{Strominger:1985ks}--\cite{Anderson:2010tc}. 
The task of computing the physical Yukawa couplings for such models can be split up into two steps: the 
calculation of the holomorphic Yukawa couplings, that is, the couplings in the superpotential, and the calculation 
of the matter field K\"ahler potential. The former relates to a holomorphic quantity and can, therefore, to some 
extent be carried out algebraically, as explained in Refs.~\cite{Candelas:1987se,Anderson:2009ge}. 
However, the matter field K\"ahler potential is non-holomorphic and its algebraic calculation does not seem to be possible - rather, 
it is likely that methods of differential geometry have to be used~\footnote{See Refs.~\cite{delaOssa:2015maa, 
Candelas:2016usb, McOrist:2016cfl} for recent progress in this direction.}. 
At present the matter field K\"ahler potential has not been worked out explicitly for any case other than the standard 
embedding (where it can be expressed in terms of the K\"ahler and complex structure moduli space metrics).

In Ref.~\cite{Blesneag:2015pvz}, we have presented a new approach to calculating the holomorphic Yukawa couplings, based entirely on methods of differential geometry. This approach was developed in the context of the simplest class of Calabi-Yau manifolds - hypersurfaces in products of projective spaces and the tetra-quadric manifold in a product of four $\mathbb{P}^1$'s in particular - and for bundles with Abelian structure groups. In its original form, as presented in Ref.~\cite{Blesneag:2015pvz}, 
this method is only applicable to a handful of Calabi-Yau manifolds. 
The purpose of the present paper is to present a significant generalisation to all complete intersection Calabi-Yau manifolds (CICY manifolds in short). 
Hence, we will show that our approach is not restricted to specific manifolds but can, in fact, 
be applied to large classes, in this case to the almost 8000 CICY manifolds classified in Refs.~\cite{Candelas:1987kf,Candelas:1987du} 
as well as to their quotients~\cite{Braun:2010vc}.  We would also like to relate our method to the earlier algebraic one~\cite{Candelas:1987se,Anderson:2009ge} and 
demonstrate that the two approaches are equivalent. Although, in the present paper, we will only discuss the holomorphic 
Yukawa couplings, we hope that the insight gained in this context will ultimately also be of use for the calculation of the matter 
field K\"ahler potential and the physical Yukawa couplings. 

In general, the perturbative, holomorphic Yukawa couplings for a line bundle model on a Calabi-Yau manifold $X$ are given by
\be 
\l (\nu_1, \nu_2, \nu_3)= \int_X \Omega \wedge \nu_1 \wedge \nu_2 \wedge \nu_3\,.
\label{holyuk}
\ee
Here, $\Omega$ is the holomorphic $(3,0)$--form on $X$ and $\nu_i\in H^1(X,K_i)$ are closed $(0, 1)$-forms, taking values in certain line bundles $K_i$ on $X$, which represent the three types of matter multiplets involved in the corresponding superpotential term. Consistency of Eq.~\eqref{holyuk} requires that $K_1 \otimes K_2 \otimes K_3= {\cal O}_X$,  where ${\cal O}_X$ is the trivial bundle on $X$. Strictly speaking, Eq.~\eqref{holyuk} needs to be evaluated for the harmonic representatives (relative to the Ricci-flat Calabi-Yau metric) of the cohomologies $H^1(X,K_i)$. Fortunately, the integral~\eqref{holyuk} is invariant under changes $\nu_i\rightarrow\nu_i+\bar\pt\x_i$ by exact forms and for this reason any representatives of $H^1(X,K_i)$ can be used.

The CICY manifolds are defined as complete intersections in ambient spaces of the form $\cA=\mathbb{P}^{n_1}\times\cdots\times\mathbb{P}^{n_m}$. 
Provided the line bundles $K_i$ are obtained as restrictions of ambient space line bundles $\cK_i\rightarrow\cA$ to $X$, we will show that the $(0,1)$-forms $\nu_i$ can be obtained from certain forms on the ambient space $\cA$ and that the integral~\eqref{holyuk} can be evaluated explicitly by converting it to an integral over the ambient space.

More precisely, we find that a closed $(0,1)$-form $\nu_i$ is, in general, related to an entire chain of ambient space $(0,a)$-forms, $\hat{\nu}_{i,a}$, where $a=1,\ldots ,k+1$ and $k$ is the co-dimension of $X$ in $\cA$. The integral~\eqref{holyuk} can then be re-written as an integral over $\cA$ which, in general, involves all forms $\hat{\nu}_{i,a}$. For a given $\nu_i$, the associated chain may terminate early, in the sense that, for a certain $\tau_i$, we have $\hat{\nu}_{i,\tau_i}\neq 0$ and $\hat{\nu}_{i,a}=0$ for all $a>\tau_i$. In this case we say that $\nu_i$ is of type $\tau_i$. One of our most important results is the vanishing theorem
\begin{equation}
 \tau_1+\tau_2+\tau_3<{\rm dim}(\cA)\quad\Longrightarrow\quad \l (\nu_1, \nu_2, \nu_3)=0\; . \label{vanth}
\end{equation} 
Particularly for high co-dimension and corresponding large ambient space dimension ${\rm dim}(\cA)$  this statement implies the vanishing of many Yukawa couplings, since cases with large types $\tau_i$ are relatively rare. The vanishing due to this theorem can not be explained by an obvious symmetry of the effective four-dimensional theory and is topological in nature. 

The outline of the paper is as follows. In the next section, we review the results of Ref.~\cite{Blesneag:2015pvz} for Calabi-Yau hypersurfaces in products of projective spaces, in order to set the scene. In Section~\ref{codim2}, we generalise to co-dimension two CICYs and in Section~\ref{higher} we deal with the general case of arbitrary co-dimension. In Section~\ref{examples}, our method is illustrated with several explicit examples and we conclude in Section~\ref{conclusion}. A number of technical issues have been moved to the appendices. Of particular importance is Appendix~\ref{appC} which explains the multiplication of harmonic forms on $\mathbb{P}^n$, 
the key ingredient required to relate our approach to the earlier algebraic methods~\cite{Candelas:1987se,Anderson:2009ge} for calculating holomorphic Yukawa couplings. 


\section{Review of Yukawa couplings for co-dimension one CICYs}
\label{review}


In this section, we will review results from Ref.~\cite{Blesneag:2015pvz} for holomorphic Yukawa couplings  on co-dimension one CICYs. 
The ambient space ${\cal A}$ consists of a product of projective factors
\be 
{\cal A}= {\mathbb P}^{n_1} \times {\mathbb P}^{n_2} \times \dots  {\mathbb P}^{n_m}\,, 
\label{Acodim1}
\ee
where $n_1 + n_2+ \dots +n_m=4$, and  the homogeneous coordinates on each $\mathbb{P}^{n_i}$ factor are denoted by $(x_i^\alpha)$, where $\alpha=0,1,\ldots,n_i$.
The Calabi-Yau three-fold $X$ is given by the zero locus of a single homogeneous polynomial $p$ in ${\cal A}$  and, 
for $X$ to be a Calabi-Yau manifold, $p$ has to have multi-degree ${\bf q}=(n_1+1, n_2+1, \dots, n_m+1)$. This defining polynomial $p$ can also be thought of as a 
global holomorphic section of the line bundle 
\be 
{\cal N}= {\cal O}_{{\cal A}} ({\bf q})\,. 
\label{Ncodim1}
\ee
On $X$, we consider a vector bundle $V$ (in one of the $E_8$ sectors), given by the sum of line bundles
\begin{equation}
 V=\bigoplus_{a=1}^nL_a\; . \label{Vdef}
\end{equation} 
We are interested in a Yukawa coupling between three matter multiplets, each associated to a closed, bundle-valued $(0,1)$-form $\nu_i$ representing a cohomology in $H^1(X,K_i)$. Here, the $K_i$ are certain line bundles on $X$ given either by the line bundles $L_a$ in Eq.~\eqref{Vdef} or their duals or tensor powers. The precise correspondence between four-dimensional matter multiplets and associated line bundles is provided in Ref.~\cite{Anderson:2012yf}. Further, we assume that the line bundles $K_i$ are obtained as restrictions of corresponding ambient space line bundles $\cK_i$, so $K_i=\cK_i|_X$ and  that, likewise, the $(0,1)$-forms $\nu_i$ are obtained as restrictions to $X$ of ambient space counterparts $\hat{\nu}_i$. Further, we require the holomorphic $(3,0)$-form $\Omega$ on $X$ as well as its ambient space counterpart $\hat{\Omega}$. With this preparation, we can attempt to evaluate Eq.~\eqref{holyuk}, by inserting an appropriate delta-function current~\cite{Candelas:1987se} in order to convert the RHS into the ambient space integral
\be 
\l (\nu_1, \nu_2, \nu_3)=-\frac{1}{2 i} \int_{{\cal A}} \hat{\Omega} \wedge \hat{\nu}_1 \wedge \hat{\nu}_2 \wedge \hat{\nu}_3
\wedge \delta^2 (p) d p \wedge d \bar p\;. 
\label{holyuk1}
\ee
To further simplify this integral, we introduce the ambient space differential forms
\begin{equation} 
\mu_j = \frac{1}{n_j!} \eps_{\a_0 \a_1 \dots \a_{n_j}} x_j^{\a_o} d x_j^{\a_1} \wedge \dots \wedge d x_j^{\a_{n_j}}\; ,\qquad\qquad
\mu =\mu_1 \wedge \mu_2 \wedge \dots \wedge \mu_m\,,  \label{mudef}
\end{equation}
and use the identities~\cite{Candelas:1987se, Strominger:1985it, Witten:1985xc, Candelas:1987kf}  
\be 
\hat{\Omega}\wedge d p = \mu\;,\qquad\qquad  \delta^2 (p) d \bar p = \frac{1}{\pi} \bar \partial \Big( \frac{1}{p} \Big)\; .
\ee
Eq.~\eqref{holyuk1} can then be converted into
\be 
\l(\nu_1,\nu_2,\nu_3) =- \frac{1}{2 \pi i}\int_{{\cal A}} \frac{\mu}{p} \wedge \Big[ {\bar \pt} \hat{\nu}_1 \wedge \hat{\nu}_2 \wedge \hat{\nu}_3-
 \hat{\nu}_1 \wedge {\bar \pt} \hat{\nu}_2 \wedge \hat{\nu}_3 +\hat{\nu}_1 \wedge  \hat{\nu}_2 \wedge {\bar \pt} \hat{\nu}_3
 \Big]\,. 
 \label{holyuk2}
 \ee
In deriving this expression, we have performed an integration by parts and ignored the boundary term. In Ref.~\cite{Blesneag:2015pvz} we have shown 
that this boundary term indeed vanishes for the ambient space ${\cal A}= {\mathbb P}^1 \times  {\mathbb P}^1 \times {\mathbb P}^1 \times {\mathbb P}^1 $
and in Appendix A we  generalise this proof to ambient spaces of the form~\eqref{Acodim1}.

It is important to note that even though the forms $\nu_i$ are closed on $X$, the forms $\hat{\nu}_i$ are not always closed on ${\cal A}$ and, hence, the integral~\eqref{holyuk2} does not necessarily vanish. To discuss this in more detail, let us focus on a line bundle $K\rightarrow X$ (which represents any of the line bundles $K_i$ above) which is obtained as a restriction $K=\cK|_X$ of a line bundle $\cK\rightarrow\cA$. For a closed $(0,1)$-form $\nu\in H^1(X,K)$ we would like to construct its ambient space counterpart $\hat{\nu} \in \Omega^1 ({\cal A}, {\cal K})$ such that $\nu =\hat{\nu}|_X$. To do this, we need to consider the relation between $K$ and $\cK$ which is governed by the Koszul short exact sequence
\be 
0 \longrightarrow {\cal N}^* \otimes {\cal K} \stackrel{p}{\longrightarrow} {\cal K}  \stackrel{r}{\longrightarrow} K  \longrightarrow 0\;, 
\label{2.9}
\ee
where ${\cal N}$ is the line bundle~\eqref{Ncodim1}, the map $p$ is multiplication by the defining polynomial $p$
and $r$ is the restriction map. This short exact sequence leads to an associated long exact sequence of cohomology groups
whose relevant part is given by
\bea
 \cdots&\longrightarrow &  H^1 ({\cal A}, {\cal N}^* \otimes {\cal K}) \stackrel{p}{\longrightarrow} H^1 ({\cal A}, {\cal K}) \stackrel{r}{\longrightarrow} H^1 (X, K)\nonumber \\
&\stackrel{\delta}{\longrightarrow} &H^2 ({\cal A}, {\cal N}^* \otimes {\cal K}) \stackrel{p}{\longrightarrow} H^2 ({\cal A}, {\cal K}) \stackrel{r}{\longrightarrow} H^2 (X, K) 
\longrightarrow \dots\;.
\label{2.10}
\eea
The map $\delta$ is called the co-boundary map. This long exact sequence allows us to write down a general expression for 
$H^1 (X, K)$, namely
\be 
H^1 (X, K) =  r \Big( {\rm Coker} \Big( H^1 ({\cal A}, {\cal N}^* \otimes {\cal K}) \stackrel{p}{\rightarrow}
H^1 ({\cal A}, {\cal K})\Big) \Big) \oplus 
\d^{-1} \Big( {\rm Ker} \Big( H^2 ({\cal A}, {\cal N}^* \otimes {\cal K}) \stackrel{p}{\rightarrow}
H^2 ({\cal A}, {\cal K})\Big) \Big) \; .
\label{codim1H1}
\ee
We see that $H^1 (X, K)$ receives two contributions, one from $H^1({\cal A}, {\cal K})$ (modulo identifications) and the 
other from (the kernel in) $H^2({\cal A}, {\cal N}^*\otimes{\cal K})$.
Let us discuss these two contributions separately, keeping in mind that the general case is a sum of these.\\[2mm]
{\bf Type 1}: 
If $\nu$ descends from $H^1({\cal A}, {\cal K})$ we refer to it as ``type 1". In this case we have a 
$(0,1)$-form $\hat{\nu}\in H^1({\cal A}, {\cal K})$ which, under the map $r$, restricts to $\nu\in H^1(X, K)$. 
Since $\hat{\nu}$ represents an ambient space cohomology it is closed, so
\begin{equation}
\bar{\partial}\hat{\nu}=0\;.
\label{2.12}
\end{equation} 
{\bf Type 2:} If $\nu$ descends from $H^2({\cal A}, {\cal N}^*\otimes{\cal K})$  we refer to it as ``type 2".
This case is more complicated since it involves the co-boundary map $\delta$ in~\eqref{2.10}. 
Following the discussion of co-boundary maps in Appendix~\ref{appB}, we can start with  
an ambient space $(0,2)$-form $\hat{\omega}=\delta(\nu) \in H^2({\cal A}, {\cal N}^*\otimes{\cal K})$ 
which is the image of $\nu$ under the co-boundary map. Then the form $\nu$ can be obtained as the restriction to
 $X$ of an ambient space $(0,1)$-form $\hat{\nu}$ which is related to $\hat{\omega}$ by the following simple equation
\be
{\bar \partial} \hat{\nu}=p \hat{\omega}  \,. 
\label{coboundmap}
\ee
Unlike in the previous case the form $\hat{\nu}$ is no longer closed. However its restriction to $X$ is closed 
because $p=0$ on $X$.\\[2mm] 
The Yukawa coupling~\eqref{holyuk} involves three $(0,1)$-forms, $\nu_1$, $\nu_2$ and $\nu_3$, each of which can be either of type 1 or type 2 (or a linear combination of both types). The simplest possibility arises when all three forms are of type 1, so that $\bar{\partial}\hat{\nu}_i=0$ for $i=1,2,3$. Then, Eq.~\eqref{holyuk2} shows that the Yukawa coupling vanishes,
\begin{equation}
 \l(\nu_1,\nu_2,\nu_3)=0\;.
 \label{2.14}
\end{equation} 
This vanishing is quasi-topological and related to the cohomology structure for 
$K_1$, $K_2$ and $K_3$ in the sequence~\eqref{2.10} - there is no expectation that it can be explained in 
terms of a symmetry in the four-dimensional theory. This is the simplest case of the vanishing theorem mentioned in the introduction.

From Eq.~\eqref{holyuk2} and using~\eqref{coboundmap} the general formula for the Yukawa coupling reads
\be 
\l(\nu_1,\nu_2,\nu_3) =- \frac{1}{2 \pi i}\int_{{\cal A}} \mu \wedge \Big[ \hat{\o}_1 \wedge \hat{\nu}_2 \wedge \hat{\nu}_3-
 \hat{\nu}_1 \wedge \hat{\o}_2 \wedge \hat{\nu}_3 +\hat{\nu}_1 \wedge  \hat{\nu}_2 \wedge  \hat{\o}_3
 \Big]\,. 
 \label{holyuk3}
 \ee
If a particular $\nu_i$ is of type 1 then the associated $\hat{\omega}_i$ is zero and the corresponding term in the above expression vanishes. In this case, $\hat{\nu}_i$ 
represents an element of the ambient space cohomology $H^1(\cA,\cK_i)$ which can be constructed explicitly as a $(0,1)$ differential form. For a $\nu_i$ of type 
two the associated $(0,2)$-form $\hat{\omega}_i$ represents an element of the ambient space cohomology $H^2(\cA,\cN^*\otimes\cK)$ and can be written 
down as a differential form, while the associated form $\hat{\nu}_i$ can be obtained by solving Eq.~\eqref{coboundmap}. In this way, all 
differential forms in the integral~\eqref{holyuk3} are known and the integral can be evaluated explicitly. Since holomorphic Yukawa couplings depend only on the cohomology 
classes of the closed forms we can take the closed  $(0,1)$ and  $(0,2)$ forms on the ambient space to be harmonic with respect to the 
Fubini-Study metric. In Ref.~\cite{Blesneag:2015pvz}, the relevant differential forms on $\mathbb{P}^1$ have been constructed and in Appendix~\ref{appC} this construction is extended to $\mathbb{P}^n$.

A simple case arises when two forms, say $\nu_1$ and $\nu_2$, are of type 1 and the remaining one, $\nu_3$, is of type 2. In this case, Eq.~\eqref{holyuk3} simplifies to
\be 
\l(\nu_1,\nu_2,\nu_3) =- \frac{1}{2 \pi i}\int_{{\cal A}} \mu \wedge 
 \hat{\nu}_1 \wedge  \hat{\nu}_2 \wedge  \hat{\o}_3\,. 
 \label{holyuk33}
 \ee
This expression involves only closed forms on the ambient space and there is no need to solve the co-boundary map~\eqref{coboundmap} for any of the three forms. In Ref.~\cite{Blesneag:2015pvz} the precise relation between the integral~\eqref{holyuk33} and the earlier algebraic calculations~\cite{Candelas:1987se,Anderson:2009ge} has been established. The main result of this discussion is that both methods - direct evaluation of the integral~\eqref{holyuk33} and the algebraic 
method - are consistent and only differ by an  overall constant which has been quantified.


\section{Yukawa couplings for co-dimension two CICYs}
\label{codim2}


In the remainder of the paper we will generalise the results reviewed in the previous section to higher co-dimension CICYs starting, in this section, with the co-dimension two case.


\subsection{Lifting forms to the ambient space}


As before, the ambient space is given by a product of projective spaces
\be  
{\cal A}= {\mathbb P}^{n_1} \times {\mathbb P}^{n_2} \times \dots  {\mathbb P}^{n_m}\; , 
\label{3.1}
\ee
but now we require that $n_1+\cdots +n_m=5$. The Calabi-Yau manifold $X$ is given by the 
common zero locus of two polynomials $p=(p_1,p_2)$ with multi-degrees ${\bf q}_1=(q_1^1,\ldots ,q_1^m)$ and ${\bf q}_2=(q_2^1,\ldots ,q_2^m)$, 
respectively. The Calabi-Yau condition, $c_1(X)=0$, translates into
\begin{equation}
 q_1^r+q_2^r=n_r+1
\end{equation}
for all $r=1,\ldots ,m$. We can also view $p$ as a global, holomorphic section of the bundle
\begin{equation}
\cN=\cO_\cA({\bf q}_1)\oplus\cO_\cA({\bf q}_2)\; .
\end{equation}
As before, we would like to understand the relation between closed line-bundle valued  $(0,1)$-forms on $X$ and certain 
forms on the ambient space $\cA$. We start with a line bundle $K\rightarrow X$, its ambient space counterpart $\cK\rightarrow\cA$ 
such that $K=\cK|_X$ and a closed $K$-valued $(0,1)$-form $\nu\in H^1(X,K)$ which represents any of the three 
forms $\nu_i$ entering the integral~\eqref{holyuk} for the holomorphic Yukawa couplings. 
The relation between $K$ and $\cK$ is still described by the Koszul sequence which, due to $X$ being defined at 
co-dimension two, is no longer short-exact but given by the four-term sequence
\be 
0 \longrightarrow \Lambda^2 {\cal N}^* \otimes {\cal K}
\stackrel{q}{\longrightarrow} {\cal N}^* \otimes {\cal K} \stackrel{p}{\longrightarrow} {\cal K}  \stackrel{r}{\longrightarrow} K  \longrightarrow 0\;.
\label{4termseq}
\ee
As before, the map $p$ acts by multiplication and $r$ is the restriction map. The map $q$ is fixed by exactness of the sequence, that is $p\circ q=0$, and by matching polynomial degrees. As a result, it is given, up to an overall, irrelevant constant, by
\be 
q =
\left( \begin{array}{rr}
-p_2 \\ p_1 
\end{array} \right)\; .
\label{3.12}
\ee
In practice, the four-term sequence~\eqref{4termseq} is best dealt with by splitting it up into the two short exact sequences
\begin{equation}
0 \longrightarrow \Lambda^2 {\cal N}^* \otimes {\cal K}
\stackrel{q}{\longrightarrow} {\cal N}^* \otimes {\cal K} \stackrel{g_1}{\longrightarrow} {\cal C}\longrightarrow 0\;,\qquad\qquad
0 \longrightarrow  {\cal C} \stackrel{g_2}{\longrightarrow} {\cal K}  \stackrel{r}{\longrightarrow} K  \longrightarrow 0\; , \label{codim2short}
\end{equation}
where ${\cal C}$ is a suitable co-kernel and $g_1$, $g_2$ are maps satisfying $g_2 \circ g_1=p$. These quantities are determined by exactness of the above two sequences but will, fortunately, not be required explicitly. The relevant parts of the two long exact sequences associated to the short exact sequences~\eqref{codim2short} read
\bea
 \cdots&\longrightarrow &  H^1 ({\cal A},  {\cal C}) \stackrel{g_2}{\longrightarrow} H^1 ({\cal A}, {\cal K}) \stackrel{r}{\longrightarrow} H^1 (X, K)\nonumber \\
&\stackrel{\delta_1}{\longrightarrow} &H^2 ({\cal A},  {\cal C}) \stackrel{g_2}{\longrightarrow} H^2 ({\cal A}, {\cal K}) 
\longrightarrow \dots\; ,
\label{codim2l1}
\eea
and
\bea
 \cdots&\longrightarrow &  H^1 ({\cal A}, \Lambda^2{\cal N}^* \otimes {\cal K}) \stackrel{q}{\longrightarrow} H^1 ({\cal A}, {\cal N}^* \otimes{\cal K}) 
 \stackrel{g_1}{\longrightarrow} H^1 ({\cal A},  {\cal C})\nonumber \\
&\stackrel{\delta_2}{\longrightarrow} &H^2 ({\cal A}, \Lambda^2{\cal N}^* \otimes {\cal K}) \stackrel{q}{\longrightarrow} 
H^2 ({\cal A}, {\cal N}^* \otimes {\cal K}) \stackrel{g_1}{\longrightarrow} H^2 ({\cal A},  {\cal C}) \nonumber \\
&\stackrel{\delta_3}{\longrightarrow} &H^3 ({\cal A}, \Lambda^2{\cal N}^* \otimes {\cal K}) \stackrel{q}{\longrightarrow} 
H^3 ({\cal A}, {\cal N}^* \otimes {\cal K}) 
\longrightarrow \dots\;.
\label{codim2l2}
\eea
Our goal is to obtain an expression for $H^1(X,K)$ in terms of ambient space cohomologies and from~\eqref{codim2l1} we find that
\be 
H^1 (X, K) =  r \Big( {\rm Coker} \Big( H^1 ({\cal A},  {\cal C}) \stackrel{g_2}{\rightarrow}
H^1 ({\cal A}, {\cal K})\Big) \Big) \oplus 
\d_1^{-1} \Big( {\rm Ker} \Big( H^2 ({\cal A}, {\cal C}) \stackrel{g_2}{\rightarrow}
H^2 ({\cal A}, {\cal K})\Big) \Big) \; .
\label{3.17}
\ee
This expression is analogous to Eq.~\eqref{codim1H1} obtained in the co-dimension one case, but here we still have to work out $H^1 ({\cal A},  {\cal C})$ and $H^2 ({\cal A},  {\cal C})$. From the second sequence~\eqref{codim2l2} they can be read off as
\bea 
H^1 ({\cal A},  {\cal C})& = & g_1 \Big( {\rm Coker} \Big( H^1 ({\cal A}, \Lambda^2{\cal N}^* \otimes {\cal K}) \stackrel{q}{\rightarrow}
H^1 ({\cal A}, {\cal N}^* \otimes {\cal K})\Big) \Big) \nonumber \\
 & \oplus & 
\d_2^{-1} \Big( {\rm Ker} \Big( H^2 ({\cal A}, \Lambda^2 {\cal N}^* \otimes {\cal K}) \stackrel{q}{\rightarrow}
H^2 ({\cal A},  {\cal N}^* \otimes {\cal K})\Big) \Big) \; , 
\label{3.18.1}
\eea
\bea 
H^2 ({\cal A},  {\cal C})& = & g_1 \Big( {\rm Coker} \Big( H^2 ({\cal A}, \Lambda^2{\cal N}^* \otimes {\cal K}) \stackrel{q}{\rightarrow}
H^2 ({\cal A}, {\cal N}^* \otimes {\cal K})\Big) \Big) \nonumber \\
 & \oplus & 
\d_3^{-1} \Big( {\rm Ker} \Big( H^3 ({\cal A}, \Lambda^2 {\cal N}^* \otimes {\cal K}) \stackrel{q}{\rightarrow}
H^3 ({\cal A},  {\cal N}^* \otimes {\cal K})\Big) \Big) \;. 
\label{3.18.2}
\eea
Substituting Eqs.~\eqref{3.18.1} and \eqref{3.18.2} into Eq.~\eqref{3.17} gives the desired formula for $H^1 (X, K)$ in terms 
of ambient space cohomology. Despite its apparent complexity, we will see that it is possible to get to a simple generalisation of the structure derived in the co-dimension one case. 

We begin by observing that $H^1 (X, K)$ receives contributions from three ambient space cohomologies, namely from $H^1 ({\cal A}, {\cal K})$, 
$H^2 ({\cal A}, {\cal N}^* \otimes {\cal K})$ and $H^3 ({\cal A}, \Lambda^2{\cal N}^* \otimes {\cal K})$ (or, more accurately, from kernels or quotients within these cohomologies).  This means that a given closed $(0,1)$-form $\nu\in H^1(X,K)$ descends, in general, from three ambient space forms, a $(0,1)$-form $\hat{\nu}$, a $(0,2)$-forms $\hat{\omega}$ and a $(0,3)$-form $\hat{\rho}$. However, a specific $\nu\in H^1(X,K)$ might not receive all three contributions. We call a $\nu\in H^1(X,K)$ ``type 1" if the associated $\hat{\omega}$ and $\hat{\rho}$ vanish and, hence, if it is determined by the $(0,1)$-form $\hat{\nu}$ only. Likewise, $\nu\in H^1(X,K)$  is called ``type 2" if the associated $\hat{\rho}$ vanishes and it is determined by the $(0,2)$-form $\hat{\omega}$. If $\nu\in H^1(X,K)$ is determined by $\hat{\rho}$ it is called ``type 3". In general, a $\nu\in H^1(X,K)$ is a linear combination of these three types but the discussion is much simplified if we focus on each type separately. In fact, it is always possible to choose of basis of $H^1(X,K)$ such that every basis element has a definite type. Let us now be more precise and discuss each of these three types in turn.\\[2mm]
{\bf Type 1}: We will refer to $\nu\in H^1(X,K)$ as ``type 1" if it descends from $H^1(\cA,\cK)$, that is, if there is a $(0,1)$-form $\hat{\nu}\in H^1(\cA,\cK)$ on the ambient space with
\be
\begin{array}{lll}
\nu =\hat{\nu}|_X&& \nu\in H^1(X,K)\\
\bar \pt \hat{\nu} =0&&\hat{\nu}\in H^1(\cA,\cK)\; .
\end{array}
\ee\\[2mm]
{\bf Type 2}: We will refer to $\nu\in H^1(X,K)$ as ``type 2" if it descends from a closed $(0,2)$ form $\hat{\omega}\in H^2(\cA,\cN^*\otimes\cK)$. To understand the relation between $\nu$ and $\hat{\omega}$ we need to chase through Eqs.~\eqref{3.17} and \eqref{3.18.2}. Starting with Eq.~\eqref{3.17} and setting $\hat{\gamma}=\delta_1(\nu)\in H^2(\cA,{\cal C})$ we know from the definition of the co-boundary map $\delta_1$ (see Appendix~\ref{appB} for a review) that there is $(0,1)$-form $\hat{\nu} \in \Omega^1 ({\cal A}, {\cal K})$ such that
\be 
\bar \pt \hat{\nu}= g_2 \hat{\gamma}\,, \quad \nu = \hat{\nu}|_X\; . 
\ee
Further, from Eq.~\eqref{3.18.2}, there is a $\hat{\omega}\in H^2(\cA,\cN^*\otimes\cK)$ with
\begin{equation}
\hat{ \gamma}=g_1\hat{\omega}\; .
\end{equation} 
Combining these last two equations, together with $g_2 \circ g_1=p$ then leads to
\be 
\bar \pt \hat{\nu}=  (g_2 \circ g_1) \hat{\o}= p \hat{\o}\,. 
\label{3.31}
\ee
To summarise this discussion, we can write down the following chain of equations
\begin{equation}
 \begin{array}{lll}
  \nu =\hat{\nu}|_X&& \nu\in H^1(X,K)\\
  \bar \pt \hat{\nu}=p \hat{\o}&& \hat{\nu}\in\Omega^1({\cal A}, {\cal K})\\
  \bar\pt\hat{\omega}=0&&\hat{\omega}\in H^2(\cA,\cN^*\otimes\cK)
\end{array}
\end{equation}  
which describes the relation between $\nu$ and the $(0,2)$-form $\hat{\omega}$ from which it descends.\\[2mm]
{\bf Type 3}: We will refer to $\nu$ as  ``type 3" if it descends from a closed $(0,3)$-form $\hat{\rho}\in H^3(\cA,\wedge^2\cN^*\otimes\cK)$ and we need to understand the relation between $\nu$ and $\hat{\rho}$. As in the case of type 2, we start with Eq.~\eqref{3.17} and define $\hat{\gamma}=\delta_1(\nu)\in H^2(\cA,{\cal C})$ and a $(0,1)$-form $\hat{\nu} \in \Omega^1 ({\cal A}, {\cal K})$ such that
\be 
\bar \pt \hat{\nu}= g_2 \hat{\gamma}\,, \quad \nu = \hat{\nu}|_X\; . 
\ee
From surjectivity of $g_1$ in the first sequence~\eqref{codim2short} we can write $\hat{\gamma}=g_1\hat{\omega}$ for an $\hat{\omega}\in \Omega^2(\cA,\cN^*\otimes\cK)$ and combining this with the previous equation leads to
\be 
\bar \pt \hat{\nu} = p \hat{\o}\,, 
\label{3.36}
\ee
as in the type 2 case. However, unlike for the type 2 case, $\hat{\o}$ is no longer closed and we need to carry out one more step. To this end, we consider Eq.~\eqref{3.18.2} and define the closed $(0,3)$-form $\hat{\rho}=\delta_3(\hat{\gamma}) \in H^3 ({\cal A}, \Lambda^2{\cal N}^* \otimes {\cal K})$. Writing out the co-boundary map $\delta_3$ (see Appendix B) now leads to
\be 
\bar \pt \hat{\o} = q \hat{\rho}\,, \quad \bar \pt \hat{\rho} =0\, .
\label{3.37}
\ee
Altogether, this gives the following chain of equations
\begin{equation}
\begin{array}{lll}
\nu =\hat{\nu}|_X&& \nu \in H^1(X, K) \\
{\bar \pt} \hat{\nu} = p  \hat{\omega}&&\hat{\nu} \in \Omega^1 ({\cal A}, {\cal K})\\
\bar \pt \hat{\o}   = q \hat{\rho}&&\hat{\o} \in \Omega^2 ({\cal A}, {\cal N}^* \otimes {\cal K})\\
\bar \pt \hat{\rho} =0 &&\hat{\rho} \in H^3 ({\cal A}, \Lambda^{2} {\cal N}^* \otimes {\cal K})
\end{array} \label{codim2desc}
\end{equation}
which describes the relation between $\nu$ and the $(0,3)$-form $\hat{\rho}$ from which it descends. \\[2mm]
In fact, the system of equations~\eqref{codim2desc} describes the general relationship between $\nu$ and the three ambient space forms $\hat{\nu}$, $\hat{\omega}$ and $\hat{\rho}$. For a given $\nu$, solving the equations~\eqref{codim2desc} gives the associated ambient space forms which, in general, are all non-zero. The three types discussed above arise from Eq.~\eqref{codim2desc} as special cases. If $\hat{\omega}=\hat{\rho}=0$ for a given $\nu$, then $\hat{\nu}$ is closed and $\nu$ is of type 1. If $\hat{\omega}\neq 0$ but $\hat{\rho}=0$ (and $\hat{\nu}$ does not have a closed part which would correspond to a type 1 component) then $\hat{\omega}$ is closed and $\nu$ is of type 2. Finally, if $\hat{\rho}\neq 0$ (and $\hat{\nu}$, $\hat{\omega}$ do not have closed parts which would correspond to type 1 and type 2 components, respectively) then $\nu$ is of type 3.

Let us point out that, in general, the set of all forms $\hat{\nu}, \ \hat{\o}, \ \hat{\rho}$ is not always identified with the entire spaces in the second column of~\eqref{codim2desc} but, rather, with kernels and co-kernels of the maps $p$ and $q$ within those spaces. In each particular case, these kernels and co-kernels can be found from Eqs.~\eqref{3.17}, \eqref{3.18.1} and~\eqref{3.18.2}.

Our goal now is to express the Yukawa couplings~\eqref{holyuk} in terms of the ambient space forms $\hat{\nu}$, $\hat{\omega}$ and $\hat{\rho}$. If $\nu$ is of a specific type, the highest non-vanishing form which appears in the Eqs.~\eqref{codim2desc} represents an ambient space cohomology and can be written down explicitly, following the rules explained in Appendix~\ref{appC}. The lower-degree forms then have to be obtained by solving the Eqs.~\eqref{codim2desc}. In this way, all relevant ambient space forms can be calculated explicitly. 


\subsection{A derivation of Yukawa couplings}
\label{derivation}


We will now derive the formula for the Yukawa couplings~\eqref{holyuk} in terms of ambient space forms. For each $(0,1)$-form $\nu_i\in H^1(X,K_i)$ involved we have an associated chain of ambient space forms $\hat{\nu}_i$, $\hat{\omega}_i$ and $\hat{\rho}_i$, in line with the Eqs.~\eqref{codim2desc}. The forms $\hat{\omega}_i$ take values in the rank two line bundle sum ${\cal N}^*\otimes\cK_i=\cO_\cA(-{\bf q}_1)\otimes\cK_i\oplus \cO_\cA(-{\bf q}_2)\otimes\cK_i$ and we denote the two corresponding components by $\hat{\omega}_i^a$, where $a=1,2$. Starting with Eq.~\eqref{holyuk}, we insert two delta-function currents
\be 
\l (\nu_1, \nu_2, \nu_3)= \frac{1}{(2 \pi i)^2} \int_{{\cal A}} \hat{\Omega} \wedge 
\hat{\nu}_1 \wedge \hat{\nu}_2 \wedge \hat{\nu}_3 \wedge dp_1 \wedge {\bar \pt}
\Big( \frac{1}{p_1}\Big) \wedge dp_2 \wedge {\bar \pt}
\Big( \frac{1}{p_2}\Big) \,. 
\label{3.6}
\ee
which converts the integral to one over the ambient space. Using the standard formula (see~\cite{Candelas:1987se, Strominger:1985it, Witten:1985xc, Candelas:1987kf})
\be 
\hat{\Omega} \wedge  dp_1 \wedge dp_2 =\mu\; ,
\ee
where $\mu$ has been defined in Eq.~\eqref{mudef}, we obtain
\be 
\l (\nu_1, \nu_2, \nu_3)= \frac{1}{(2 \pi )^2} \int_{{\cal A}} \mu \wedge 
\hat{\nu}_1 \wedge \hat{\nu}_2 \wedge \hat{\nu}_3   \wedge {\bar \pt}
\Big( \frac{1}{p_1}\Big)  \wedge {\bar \pt}
\Big( \frac{1}{p_2}\Big) \,. 
\label{3.8}
\ee
Now we have to integrate by parts twice ignoring the boundary integrals which do not contribute (see Appendix A). 
After the first integration we obtain
\be 
\l(\nu_1,\nu_2,\nu_3) = \frac{1}{(2 \pi )^2}\int_{{\cal A}} \frac{\mu}{p_1} \wedge \Big[\bar \pt \hat{\nu}_1 \wedge \hat{\nu}_2 \wedge \hat{\nu}_3-
 \hat{\nu}_1 \wedge \bar \pt \hat{\nu}_2 \wedge \hat{\nu}_3 +\hat{\nu}_1 \wedge  \hat{\nu}_2 \wedge  \bar \pt \hat{\nu}_3
 \Big] \wedge {\bar \pt}
\Big( \frac{1}{p_2}\Big) \,. 
\label{codim2int2}
 \ee
 The derivatives of $\hat{\nu}_i$ can be evaluated using~\eqref{codim2desc}. This leads to
 \be 
\bar \pt \hat{\nu}_1 \wedge \hat{\nu}_2 \wedge \hat{\nu}_3-
 \hat{\nu}_1 \wedge \bar \pt \hat{\nu}_2 \wedge \hat{\nu}_3 +\hat{\nu}_1 \wedge  \hat{\nu}_2 \wedge  \bar \pt \hat{\nu}_3 := p\hat{\b} =
 p_1 \hat{\b}^{1}  +p_2 \hat{\b}^{2} \,, 
 \ee
 where $\hat{\b}$ is a vector with components given by
\be
\hat{\b}^{1} =\hat{\o}_1^{1}  \wedge \hat{\nu}_2 \wedge \hat{\nu}_3-
\hat{\nu}_1 \wedge \hat{\o}_2^{1} \wedge \hat{\nu}_3 +\hat{\nu}_1 \wedge  \hat{\nu}_2 \wedge  \hat{\o}_3^{1}\;,\quad 
\hat{\b}^{2} =\hat{\o}_1^{2}  \wedge \hat{\nu}_2 \wedge \hat{\nu}_3-
\hat{\nu}_1 \wedge \hat{\o}_2^{2} \wedge \hat{\nu}_3 +\hat{\nu}_1 \wedge  \hat{\nu}_2 \wedge  \hat{\o}_3^{2}\,.
\ee
Substituting these expressions back into the integral~\eqref{codim2int2}, we note that the term $p_2 \hat{\b}^{2}$ does not contribute
since $p_2 \bar \pt  \Big( \frac{1}{p_2}\Big) \sim p_2 \d^2 (p_2) d {\bar p}_2=0$ and that we are, hence, left with
\be 
\l(\nu_1,\nu_2,\nu_3) = \frac{1}{(2 \pi )^2}\int_{{\cal A}} \mu \wedge \hat{\b}^{1} \wedge {\bar \pt}
\Big( \frac{1}{p_2}\Big) =-
\frac{1}{(2 \pi )^2}\int_{{\cal A}} \frac{\mu }{p_2} \wedge \bar \pt \hat{\b}^{1}\,. 
 \ee
Using Eqs.~\eqref{codim2desc} it is easy to work out that $\bar \pt\hat{\b}^{1}=- p_2 \hat{\eta}$, where
\bea
\hat{\eta} &=& \hat{\rho}_1 \wedge \hat{\nu}_2 \wedge \hat{\nu}_3 + \hat{\nu}_1 \wedge \hat{\rho}_2 \wedge \hat{\nu}_3+
\hat{\nu}_1 \wedge \hat{\nu}_2 \wedge \hat{\rho}_3+ \hat{\nu}_1 \wedge \hat{\o}_2^{2} \wedge \hat{\o}_3^{1}-  \hat{\nu}_1 \wedge \hat{\o}_2^{1} \wedge \hat{\o}_3^{2}
\nonumber \\
&&+   \hat{\o}_1^{1} \wedge \hat{\nu}_2 \wedge  \hat{\o}_3^{2}  -   \hat{\o}_1^{2} \wedge \hat{\nu}_2 \wedge  \hat{\o}_3^{1}
+   \hat{\o}_1^{2} \wedge  \hat{\o}_2^{1}  \wedge \hat{\nu}_3-    \hat{\o}_1^{1} \wedge  \hat{\o}_2^{2}  \wedge \hat{\nu}_3\,.
 \label{etadef}
\eea
Hence, the final expression for the Yukawa coupling is 
\be 
\l(\nu_1,\nu_2,\nu_3) =
\frac{1}{(2 \pi )^2}\int_{{\cal A}} \mu  \wedge \hat{\eta}
 \label{codim2res}
 \ee
with $\hat{\eta}$ given in~\eqref{etadef}. Eq.~\eqref{codim2res} together with Eq.~\eqref{etadef} is our main general result for the co-dimension two case. As we will see  in Section~\ref{examples}, this result, together with the expressions for ambient space harmonic forms in Appendix~\ref{appC} and Eq.~\eqref{codim2desc}, allows for an explicit calculation of the holomorphic Yukawa couplings. 

It is worth discussing a number of special cases. If all three forms $\nu_i$ are of type 1 then $\hat{\omega}_i=\hat{\rho}_i=0$, for $i=1,2,3$ and as a results $\hat{\eta}$ in Eq.~\eqref{etadef} and, hence, the Yukawa coupling vanishes. Now suppose two of the forms $\nu_i$, say $\nu_1$ and $\nu_2$ are of type 1, while $\nu_3$ is of type 2. In this case we have $\hat{\omega}_i=\hat{\rho}_i=0$ for $i=1,2$ and $\hat{\rho}_3=0$ so that $\hat{\eta}$ in Eq.~\eqref{etadef} and the Yukawa coupling still vanishes. These observations can be summarised by the following\\[2mm]
{\bf Theorem}: Assume that the forms $\nu_i$ which enter the integral~\eqref{holyuk} for the Yukawa couplings are of type $\tau_i$, where $i=1, 2, 3$. Then
\be 
\tau_1+ \tau_2 +\tau_3 < {\rm dim}( {\cal A}) =5\qquad\Longrightarrow\qquad \l (\nu_1,\nu_2,\nu_3) =0\; . \label{vanth2}
\ee\\[2mm]
For co-dimension one we have observed that the Yukawa coupling vanishes if all three forms $\nu_i$ are of type 1. The above vanishing theorem generalises this statement to the case of co-dimension two. 

There are two special cases for which the expression~\eqref{codim2res} simplifies considerably. Firstly, assume that the types of the $(0,1)$-forms $\nu_i$ are given by $(\tau_1,\tau_2,\tau_3)=(1,1,3)$. Then we have from Eqs.~\eqref{codim2res} and \eqref{etadef}
\begin{equation}
\l(\nu_1,\nu_2,\nu_3) =
\frac{1}{(2 \pi )^2}\int_{{\cal A}} \mu  \wedge \hat{\nu}_1 \wedge \hat{\nu}_2 \wedge \hat{\rho}_3\; , \label{yuk113}
\end{equation}
and all three bundle-valued forms in the integrand represent ambient space cohomologies. The other simple case arises for types $(\tau_1,\tau_2,\tau_3)=(1,2,2)$  where Eq.~\eqref{codim2res} becomes
\be
 \l(\nu_1,\nu_2,\nu_3) =
\frac{1}{(2 \pi )^2}\int_{{\cal A}} \mu  \wedge\hat{\nu}_1\wedge\hat{\omega}_2\wedge\hat{\omega}_3\; , \label{yuk122}
\ee
with an anti-symmetric contraction of the bundle indices for $\hat{\omega}_i$ understood. Again, all three forms in the integrand represent ambient space cohomologies. 

We will now proceed to arbitrary co-dimension and show that analogous statements can be obtained in the general case. 


\section{Generalisation to higher co-dimensions}
\label{higher}


\subsection{Lifting forms to the ambient space}


We will now tackle the case of arbitrary co-dimension starting, as before, with the problem of writing closed line bundle-valued $(0,1)$-forms on the Calabi-Yau manifold in terms of ambient space forms. Our ambient space remains the product of projective spaces
\be 
{\cal A}= {\mathbb P}^{n_1} \times {\mathbb P}^{n_2} \times \dots  {\mathbb P}^{n_m}\; ,
\label{4.1}
\ee
where now $n_1+\cdots +n_m=3+k$ and $k$ is the co-dimension. The CICY manifold $X\subset\cA$ is defined as the common zero locus of $k$ homogeneous polynomials $p_a$ with multi-degrees ${\bf q}_a=(q_a^1,\ldots ,q_a^m)$, where $a=1,\ldots , k$. The Calabi-Yau condition, $c_1(X)=0$ now reads
\begin{equation}
 \sum_{a=1}^kq_a^r=n_r+1
\end{equation}
for all $r=1,\ldots ,m$. As before, we combine these polynomials into the row vector $p=(p_1,\ldots ,p_k)$ which can be viewed as a section of the line bundle sum
\begin{equation}
 \cN=\cO_\cA({\bf q}_1)\oplus\cdots\oplus \cO_\cA({\bf q}_k)\; .
\end{equation} 
The relation between a line bundle $K\rightarrow X$ and its ambient space counterpart $\cK\rightarrow\cA$ (such that $K=\cK|_X$) is again governed by the Koszul sequence
\be 
0 \longrightarrow \Lambda^k {\cal N}^* \otimes {\cal K}  \stackrel{q_{k}}{\longrightarrow} 
 \Lambda^{k-1} {\cal N}^* \otimes {\cal K}  \stackrel{q_{k-1}}{\longrightarrow}   \dots 
 \stackrel{q_{2}}{\longrightarrow}
{\cal N}^* \otimes {\cal K} \stackrel{q_1=p}{\longrightarrow} {\cal K}  \stackrel{q_0=r}{\longrightarrow} K  \longrightarrow 0\;, 
\label{kosgen}
\ee
which now consists of $k+2$ terms and contains maps $q_a$ satisfying $q_a \circ q_{a+1} =0$ for all $a=0,\ldots ,k-1$. As previously, $q_0=r$ is the restriction map, $q_1=p$ is the map acting by multiplication with the polynomial vector $p$ and the higher maps $q_a$ for $a>1$ are the obvious tensor maps induced by $p$.  An $(a+1)$--form $\hat{\nu}$ taking values in $\wedge^{a}\cN^*\otimes\cK$ has components $\hat{\nu}^{b_1\cdots b_{a}}$, with completely anti-symmetrised upper indices, and the action of $q_{a}$ on this form can be explicitly written as
\begin{equation}
\left(q_{a}\hat{\nu}\right)^{b_1\cdots b_{a-1}}=p_b\,\hat{\nu}^{b_1\cdots b_{a-1}b}\; . \label{qaction}
\end{equation}
Splitting \eqref{kosgen} up into $k$ short exact sequences and chasing through the associated long exact sequences shows that $H^1(X,K)$ can now receive contributions from the $k+1$ ambient space cohomologies $H^1(\cA,\cK),$ $H^2(\cA,\cN^*\otimes\cK),\;\ldots\,,H^k(\cA,\wedge^{k-1}\cN^*\otimes\cK),\;H^{k+1}(\cA,\wedge^k\cN^*\otimes\cK)$. A closed $K$-valued $(0,1)$-form $\nu\in H^1(X,K)$ is, therefore, related to a chain of $k+1$ ambient space $(0,a)$-forms $\hat{\nu}_a$, where $a=1,\ldots ,k+1$. The precisely relationship between $\nu$ and $\hat{\nu}_a$ can be derived by a straightforward generalisation of the co-dimension two case discussed in the previous section. The result is
\begin{equation}
\begin{array}{lllllll}
\nu &=&\hat{\nu}_1|_X&\qquad&\nu& \in& H^1(X, K)\\
{\bar \pt} \hat{\nu}_1 &=& q_1  \hat{\nu}_2&&\hat{\nu}_1& \in& \Omega^1 ({\cal A}, {\cal K})\\
\bar \pt \hat{\nu}_2  & =& q_2 \hat{\nu}_3&&\hat{\nu}_2 &\in& \Omega^2 ({\cal A}, {\cal N}^* \otimes {\cal K})\\
&\vdots&&&&\vdots&\\
\bar \pt \hat{\nu}_k   &=& q_k \hat{\nu}_{k+1}&&\hat{\nu}_{k} &\in& \Omega^k ({\cal A}, \Lambda^{m-1}{\cal N}^* \otimes {\cal K})\\
\bar \pt \hat{\nu}_{k+1} &=&0&&\hat{\nu}_{k+1} &\in& H^{k+1} ({\cal A}, \Lambda^m{\cal N}^* \otimes {\cal K}) \;.
\end{array}\label{nuresgen}
\end{equation}
Note that, just like in the co-dimension two case, the forms $\hat{\nu}_a$ should be thought of as elements of certain kernels and co-kernels of the maps $q_a$ within the spaces  on the right-hand side of Eq.~\eqref{nuresgen}. For a given $\nu\in H^1(X,K)$ the associated chain of ambient space forms is obtained by solving the above equations and, in general, this leads to $k+1$ non-trivial forms $\hat{\nu}_a$. However, as before, it is useful to introduce the type $\tau$ of $\nu$ which can now take the values $\tau\in\{1,\ldots ,k+1\}$. We say that $\nu$ is of type $\tau$ if $\hat{\nu}_\tau\neq 0$, $\hat{\nu}_a=0$ for all $a>\tau$ and all $\hat{\nu}_a$ for $a<\tau$ do not contain any $\bar\pt$-closed parts. In this case, $\nu$ descends, via the Eqs.~\eqref{nuresgen}, from the $\bar\pt$-closed $(0,\tau)$-form $\hat{\nu}_\tau$ which defines an element of $H^\tau(\cA,\wedge^{\tau-1}\cN^*\otimes\cK)$.


\subsection{The structure of Yukawa couplings and a vanishing theorem}


Each of the three forms $\nu_i\in H^1(X,K_1)$ involved in the Yukawa coupling has, from Eq.~\eqref{nuresgen}, an associated chain of ambient space forms which we denote by $\hat{\nu}_{i,a}$, where $a=1,\ldots ,k+1$. To derive the general expression for the Yukawa couplings we start with~\eqref{holyuk}, insert $k$ delta-function currents and use the standard formula (see~\cite{Candelas:1987se, Strominger:1985it, Witten:1985xc, Candelas:1987kf})
\be 
\hat{\Omega} \wedge  dp_1 \wedge \cdots\wedge dp_k =\mu\; ,
\ee
where $\mu$ has been defined in Eq.~\eqref{mudef}. This leads to
\bea
\lambda (\nu_1, \nu_2, \nu_3) &=& 
\Big(- \frac{1}{2 \pi i}\Big)^k \int_{{\cal A}} \hat{\Omega} \wedge 
\hat{\nu}_{1,1} \wedge \hat{\nu}_{2,1} \wedge \hat{\nu}_{3,1} \wedge dp_1 \wedge {\bar \pt}
\Big( \frac{1}{p_1}\Big) \wedge  \dots \wedge d p_k \wedge {\bar \pt} \Big( \frac{1}{p_k} \Big) 
\nonumber \\
& =& 
\frac{\tilde{C}_k}{(2\pi)^kk!}\eps_{b_1\cdots b_k} \int_{{\cal A}} \mu \wedge 
\hat{\nu}_{1,1} \wedge \hat{\nu}_{2,1} \wedge \hat{\nu}_{3,1}  \wedge {\bar \pt}
\Big( \frac{1}{p_{b_1}}\Big)  \wedge \dots \wedge {\bar \pt} \Big( \frac{1}{p_{b_k}} \Big) \,,
\label{yukstart}
\eea
where $\tilde{C}_k =(-1)^{k(k+1)/2}\, i^k$ is a phase factor. Integrating the first $\bar\pt$ operator by parts (ignoring the boundary terms whose vanishing 
can be shown in the same way as in Appendix A) and using Eqs.~\eqref{qaction}, \eqref{nuresgen}  this turns into
\bea
 \lambda (\nu_1, \nu_2, \nu_3) &=& \frac{\tilde{C}_k}{(2\pi)^kk!}\eps_{b_1\cdots b_k} \int_{{\cal A}} \mu \wedge \left(\hat{\nu}_{1,2}^{b_1} \wedge \hat{\nu}_{2,1} \wedge \hat{\nu}_{3,1}-\hat{\nu}_{1,1} \wedge \hat{\nu}_{2,2}^{b_1} \wedge \hat{\nu}_{3,1}+\hat{\nu}_{1,1} \wedge \hat{\nu}_{2,1} \wedge \hat{\nu}_{3,2}^{b_1}\right)\nonumber\\
 &&\qquad\qquad\qquad\qquad\; \wedge\,\bar\pt\Big( \frac{1}{p_{b_2}}\Big)  \wedge \dots \wedge {\bar \pt} \Big( \frac{1}{p_{b_k}} \Big)\; .
\eea 
Here, the relation
\be 
p_b\, {\bar \pt} \Big( \frac{1}{p_b} \Big)=0
\ee
has led to the insertion of $\delta_b^{b_1}$ from Eq.~\eqref{qaction} so that we remain with a sum over $b_1$, as indicated above (while the resulting factor $p_{b_1}$ from Eq.~\eqref{qaction} cancels against $1/p_{b_1}$). We can now continue integrating by parts until all factors of the form $\bar\pt(1/p_b)$ are used up. Each of these factors leads to a partial differentiation of all forms $\hat{\nu}_{i,a}^{b_1\cdots b_{a-1}}$ which appear in the integral, effective replacing them by the forms $\hat{\nu}_{i,a+1}^{b_1\cdots b_{a-1}b}$, which appear one step lower down in the chain~\eqref{nuresgen}. Since there are $k$ such partial integrations to be performed, starting with  three $(0,1)$-forms, the end result is a sum which contains all product of three forms whose degree sums up to ${\rm dim}(\cA)=3+k$. This leads to
\begin{equation}
  \lambda (\nu_1, \nu_2, \nu_3) = \frac{C_k}{(2\pi)^k}\sum_{\substack{a_1,a_2,a_3=1\\a_1+a_2+a_3={\rm dim}(\cA)}}^k (-1)^{s(a_1,a_2,a_3)} \int_{{\cal A}} \mu \wedge \hat{\nu}_{1,a_1}\wedge \hat{\nu}_{2,a_2}\wedge\hat{\nu}_{3,a_3}\; . \label{yukfin}
\end{equation}
where $s(a_1,a_2,a_3)=(a_1+1)a_2+a_1a_3+a_2a_3$ determines the relative signs of the terms and $C_k=(-1)^{k(k+1)/2} (-1)^{[(k+1)/2]}\, i^k$ is another phase. In this formula, the bundle indices have been suppressed so the wedge product should be understood as including an appropriate tensoring of the bundle directions to form a singlet, via anti-symmetrisation by $\eps_{b_1\cdots b_k}$. The anti-symmetrisation is achieved by summing in every case as many terms with permuted indices as required for complete anti-symmetry, each with a factor $1$ or $-1$ and no additional overall normalisation. This means that, for example, $\hat{\nu}_{1,2}\wedge \hat{\nu}_{2,2}\wedge\hat{\nu}_{3,1}=\e_{b_1b_2}\hat{\nu}_{1,2}^{b_1}\wedge \hat{\nu}_{2,2}^{b_2}\wedge\hat{\nu}_{3,1}$ while $\hat{\nu}_{1,3}\wedge \hat{\nu}_{2,1}\wedge\hat{\nu}_{3,1}=\frac{1}{2}\e_{b_1b_2}\hat{\nu}_{1,3}^{b_1b_2}\wedge \hat{\nu}_{2,1}\wedge\hat{\nu}_{3,1}$.

Eq.~\eqref{yukfin} is our main general result for the holomorphic Yukawa couplings. All the ambient space forms $\hat{\nu}_{i,a}$ can be constructed explicitly, starting with Appendix~\eqref{appC} in order to write down (harmonic) representatives for ambient space cohomology for the highest degree non-trivial forms in the chain~\eqref{nuresgen} and then solving these equations to find all associated lower-degree forms. With these forms inserted, the integral~\eqref{yukfin} can be carried out explicitly, as we will demonstrate for the examples in Section~\ref{examples}.

As before, it is useful to discuss some special cases. First assume, that the $(0,1)$-forms $\nu_i$ are of type $\tau_i$ so that $\hat{\nu}_{i,a}=0$ for all $a>\tau_i$. 
If the $\tau_i$ sum up to less than the ambient space dimension ${\rm dim}(\cA)$ then all terms in Eq.~\eqref{yukfin} vanish due to the summation constraint. As a result the Yukawa coupling vanishes. Let us formulate this concisely:\\[2mm]
{\bf Theorem}: Assume that the forms $\nu_i$ which enter the integral~\eqref{holyuk} for the Yukawa couplings are of type $\tau_i$, where $i=1, 2, 3$. Then
\be 
\tau_1+ \tau_2 +\tau_3 < {\rm dim}( {\cal A}) \qquad\Longrightarrow\qquad \l (\nu_1,\nu_2,\nu_3) =0\; . \label{vantheorem}
\ee\\[2mm]
This is the general version of the vanishing theorem we have already seen for co-dimensions one and two in previous sections. As we have discussed, the type $\tau$ of a form $\nu\in H^1(X,K)$ is determined by the cohomology $H^\tau(\cA,\wedge^{\tau-1}\cN^*\otimes\cK)$ from which it descends via successive co-boundary maps. As a rule of thumb, large $\tau$'s are relatively rare since they require many non-trivial co-boundary maps and cohomologies.  Consequently, for a large ambient space dimension ${\rm dim}(\cA)$ the condition in \eqref{vantheorem} is frequently satisfied and many Yukawa couplings vanish. We stress again that vanishing due to \eqref{vantheorem} appears to be topological in nature, that is, these couplings vanish despite being allowed by the obvious symmetries of the four-dimensional effective theory.

Another special case of interest is for types $\tau_i$ satisfying $\tau_1+\tau_2+\tau_3={\rm dim}(\cA)$. In this case, only one term in \eqref{yukfin} contributes and the integral simplifies to
\begin{equation}
 \lambda (\nu_1, \nu_2, \nu_3) \sim \frac{1}{(2\pi)^k} \int_{{\cal A}} \mu \wedge \hat{\nu}_{1,\tau_1}\wedge \hat{\nu}_{2,\tau_2}\wedge\hat{\nu}_{3,\tau_3}\; , \label{yukfinsp}
\end{equation}
where we have dropped an overall phase factor. Note that, unlike in the general case~\eqref{yukfin}, all three forms $\hat{\nu}_{i,\tau_i}$ in the integrand are closed and represent ambient space cohomologies in $H^{\tau_i}(\cA,\wedge^{\tau_i-1}\cN^*\otimes\cK)$. They can, therefore, be directly constructed from the rules given in Appendix~\ref{appC}, without any need to solve Eqs.~\eqref{nuresgen}.


\section{Examples}\label{examples}
In this section, we will illustrate our general statements for models on a certain co-dimension two CICY and show that the relevant ambient space integrals can, 
in fact, be carried out explicitly. We begin by introducing the specific CICY and its properties, then move on to describing line bundles and 
line bundle-valued forms before we derive two more specific formulae for the 
Yukawa couplings for types $(\tau_1,\tau_2,\tau_3)=(1,1,3)$ and $(\tau_1,\tau_2,\tau_3)=(1,2,2)$, respectively. 
These results are then applied to three examples, each defined by a certain line bundle sum on the relevant CICY.

\subsection{A co-dimension two CICY and its properties}
Our chosen CICY is a co-dimensional two manifold in the ambient space $\mathcal{A} = \mathbb{P}^{1} \times \mathbb{P}^{1} \times \mathbb{P}^{1} \times \mathbb{P}^{1} \times \mathbb{P}^{1}$, whose homogeneous coordinates we either denote by ${\bf x}=(x_i^\alpha)$, where $i=1,\ldots ,5$ and $\alpha=0,1$ or, more explicitly, by ${\bf x}=((x_0,x_1), (y_0,y_1),( u_0,u_1),( v_0, v_1),( w_0, w_1))$. We also introduce affine coordinates $z_i=x_i^1/x_i^0$ on the coordinate patch of $\cA$ where all $x_i^0\neq 0$. The CICY is defined as the common zero locus in $\cA$ of two homogeneous polynomials $p=(p_1,p_2)$ with multi-degrees ${\bf q}_1=(0,1,1,1,1)$ and ${\bf q}_2=(2,1,1,1,1)$, respectively. This information is often summarised by the configuration matrix
\begin{eqnarray}
X=\begin{pmatrix}
\mathbb{P}^{1}& \vline & 0 & 2  \\
\mathbb{P}^{1}&\vline & 1 & 1  \\
\mathbb{P}^{1} &\vline & 1 & 1  \\
\mathbb{P}^{1} &\vline & 1 & 1 \\
\mathbb{P}^{1}& \vline & 1 & 1
\end{pmatrix}_{- 80}^{5, 45}\label{conf7487}
\end{eqnarray}
whose columns are given by ${\bf q}_1$ and ${\bf q}_2$. Attached as a superscript are the Hodge numbers $h^{1,1}(X),\,h^{2,1}(X)$ and as a subscript the Euler number, $\eta(X)$. In the standard list of Refs.~\cite{Candelas:1987kf,Candelas:1987du}, this manifold carries the number $7487$.  The defining polynomials $p=(p_1,p_2)$ can also be viewed as a section of the line bundle sum
\begin{equation}
 \cN=\cO_\cA({\bf q}_1)\oplus\cO_\cA({\bf q}_2)\; .
\end{equation}
For later reference, we also define ${\bf q}={\bf q}_1+{\bf q}_2=(2,2,2,2,2)$ and note that
\be
 \wedge^2\cN=\cO_\cA({\bf q})\; .
\ee 
In order to reduce the size of the problem, it will frequently be useful to work on a discrete quotient of the above manifold. In fact, $X$ has a freely-acting symmetry $\Gamma = \mathbb{Z}_2 \times \mathbb{Z}_2$ whose generators act on the homogeneous coordinates as
\be
\g(g_1) = \mathbbm{1}_5\times\begin{pmatrix} 1 & 0 \\ 0 & -1 \end{pmatrix}\;,\qquad \g(g_2) = \mathbbm{1}_5\times\begin{pmatrix} 0 & 1 \\ 1 & 0 \end{pmatrix}\; . \label{Z2Z2}
\ee
while the action on the defining polynomials is
\be
 \r(g_1)={\rm diag}(1,-1)\;,\qquad \r(g_2)={\rm diag}(1,-1)\; . \label{Z2Z2pol}
\ee 
The quotient $\tilde{X}=X/\G$ is a Calabi-Yau manifold with Euler number $\eta(\tilde{X})=\eta(X)/|\G|=-20$ and Hodge numbers $h^{1,1}(\tilde{X})=5$, $h^{2,1}(\tilde{X})=15$. 

\subsection{Line bundles and line bundle-valued harmonic forms}

The CICY defined by \eqref{conf7487} is favourable, by which we mean that the entire second cohomology of $X$ descends from the ambient space. This implies that every line bundle $L\rightarrow X$ can be obtained as a restriction $L=\cL|_X$ of an ambient space line bundle $\cL=\cO_\cA({\bf l})$, where ${\bf l}=(l^1,\ldots ,l^5)$. In order to compute Yukawa integrals we need to understand the cohomology of such ambient space line bundles and write down explicit differential forms representing these cohomologies. Since we are merely dealing with a product of projective spaces this is indeed possible. In the following, we briefly review the relevant results for products of $\mathbb{P}^1$ factors obtained in Ref.~\cite{Blesneag:2015pvz}. Their generalisation to arbitrary $\mathbb{P}^n$-factors can be found in Appendix~\ref{appC}.

We begin by discussing the cohomology dimensions for a line bundle $\cL=\cO_\cA({\bf l})$ which can be obtained by combining Bott's formula for line bundle cohomology on $\mathbb{P}^1$ and the K\"unneth formula. Firstly, all cohomologies of $\cL$ vanish if at least one of the integers $l^i$ equals $-1$. If all $l^i \neq - 1$, then there is precisely one non-vanishing cohomology $H^q(\mathcal{A}, \mathcal{K})$, and $q$ equals the number of integers $l^i$ with $l^i \leq -2$. The dimension of this one non-vanishing cohomology is given by
\be 
h^q(\mathcal{A}, \mathcal{L}) = \prod_{i: l^i \geq 0} (l^i+1)\prod_{i: l^i \leq -2} (-l^i-1)\; . \label{hq}
\ee
The $\cL$-values $(0,q)$-forms representing $H^q(\cA,\cL)$ can be written down as
\be
 \a_{(\mathbf{l})}=P_{(\mathbf{l})}\prod_{i: l^i \leq -2} \kappa_i^{l^i} d \overline{z}_i \label{formsaff}
\ee
where $\k_i=1+|z_i|^2$ and $P_{(\mathbf{l})}$ is a polynomial of degree $l^i$ in $z_i$, if $l^i \geq 0$, and of degree $-l^i-2$ in $\overline{z}_i$, if $l^i \leq -2$. In fact, the above forms are harmonic (relative to the Fubini-Study metric) and are, hence, in ono-to-one correspondence with the elements of $H^q(\cA,\cL)$. In particular, note that the number of arbitrary coefficients in the polynomial $P_{(\mathbf{l})}$ equals the dimension~\eqref{hq} of the cohomology group.

The above differential forms have been written down in affine coordinates $z_i$. A useful equivalent version in terms of homogeneous coordinates is given by
\be
\a_{(\mathbf{l})}=\tilde{P}_{(\mathbf{l})}\prod_{i: l^i \leq -2} \sigma_i^{l^i} d \overline{\mu}_i,
\ee
where $\tilde{P}_{(\mathbf{l})}$ is the homogeneous counterpart of $P_{(\mathbf{l})}$ and
\be
\sigma_i = \vert x_i^0\vert^2 + \vert x_i^1\vert^2\; ,\qquad \mu_i = \epsilon_{\alpha \beta} x_i^{\alpha}x_i^{\beta}\; .
\ee

The Yukawa couplings involve wedge products of differential forms and we should, therefore, understand what happens if we form wedge products of the above forms. To be specific let us consider a form $\a_{({\bf l})}$ with associated polynomial $\tilde{P}_{({\bf l})}$, representing the cohomology $H^p(\cA,\cO_\cA({\bf l}))$ and a form $\beta_{({\bf m})}$ with associated polynomial $\tilde{Q}_{({\bf m})}$, representing the cohomology $H^q(\cA,\cO_\cA({\bf m}))$. It is clear that $\alpha_{({\bf l})}\wedge \beta_{({\bf m})}$ is $\bar\pt$--closed and represents an element of $H^{p+q}(\cA,\cO_\cA({\bf l}+{\bf m}))$, however, this will, in general not be the harmonic representative. We can ask how this harmonic representative, which we denote by $\g_{({\bf l}+{\bf m})}$ with associated  polynomial $\tilde{R}_{({\bf l}+{\bf m})}$,  can be obtained from $\alpha_{({\bf l})}$ and $\beta_{({\bf m})}$. Fortunately, there is a simple answer which can be expressed in terms of the associated polynomials $\tilde{P}_{({\bf l})}$, $\tilde{Q}_{({\bf m})}$ and $\tilde{R}_{({\bf l}+{\bf m})}$. For a product of $\mathbb{P}^1$ spaces this has been derived in Ref.~\cite{Blesneag:2015pvz}. In Appendix~\ref{appC}, we explain how harmonic forms on a single $\mathbb{P}^n$ are multiplied. These results can be easily applied to a product of projective spaces with arbitrary dimensions and lead to
\be
\tilde{R}_{(\mathbf{l+m})} = c_{\mathbf{l,m}} \tilde{P}_{(\mathbf{l})} \tilde{Q}_{(\mathbf{m})}\; , \label{RPQ}
\ee
where $c_{\mathbf{l,m}}$ is a numerical coefficient explicitly given by
\be
 c_{\mathbf{l,m}} = \prod_{i:l^i\leq -2} c_{l^i, m^i} \prod_{j:m^j\leq -2} c_{m^j, l^j}\; ,\qquad c_{l,m} = \dfrac{(-l-m-1)!}{(-l-1)!}.
\ee
The polynomial multiplication on the RHS of Eq.~\eqref{RPQ} is understood with a replacement of coordinates by associated partial derivatives whenever positive degrees meet negative degrees. More specifically, whenever coordinates $x_i^\alpha$ in $\tilde{P}_{(\mathbf{k})}$ act on coordinates $\bar{x}_i^\alpha$ in $\tilde{Q}_{(\mathbf{l})}$, the former should be replaced by $\pt/\pt\bar{x}_i^\alpha$. 

In the following, we would like to further evaluate the Yukawa couplings for our example manifold and certain specific types. We will work within our familiar setting, that is, we have three line bundles $K_i=\cO_X({\bf k}_i)$ on $X$ underlying the expression for the Yukawa couplings. These line bundles descend from their ambient space counterparts $\cK_i=\cO_\cA({\bf k}_i)$ and have to satisfy the condition
\be
K_1\otimes K_2\otimes K_3=\cO_X\quad\Longrightarrow\quad{\bf k}_1+{\bf k}_2+{\bf k}_3=0\; . \label{kzero}
\ee
We would like to calculate the Yukawa couplings for three $K_i$--valued $(0,1)$--forms $\nu_i\in H^1(X,K_i)$. From the Eqs.~\eqref{codim2desc} each of these comes with a chain of ambient space forms, namely the $(0,1)$--forms $\hat{\nu}_i$, the $(0,2)$--forms $\hat{\omega}_i$ and the $(0,3)$--forms $\hat{\rho}_i$ which enter the general formula~\eqref{codim2res} for the Yukawa couplings. In the following, we focus on certain cases where the $\nu_i$ have specific types $\tau_i$.

\subsection{Yukawa couplings of type $(1,1,3)$}
We now assume that two of the forms $\nu_i$, say $\nu_1$ and $\nu_2$ for definiteness, are of type 1, while $\nu_3$ is of type 3. Note that this saturates the bound in Eq.~\eqref{vanth2} and constitutes one of the two simplest cases for co-dimension two to which the vanishing theorem does not apply (the other one being discussed in the next sub-section). In this case, the Yukawa couplings are given by Eq.~\eqref{yuk113} which only involves the ambient space forms $\hat{\nu}^1\in H^1(\cA,\cK_1)$, $\hat{\nu}^2\in H^1(\cA,\cK_2)$ and $\hat{\rho}_3\in H^3(\cA,\wedge^2\cN^*\otimes\cK_3)$.

Following the rules for cohomology explained in the last sub-section, in order for $H^1(\cA,\cK_1)$ and $H^1(\cA,\cK_2)$ to 
be non-trivial, we require that ${\bf k}_1$ and ${\bf k}_2$  each have precisely one entry 
less than or equal to $-2$ and all other entries positive. 
Further, for $H^3(\cA,\wedge^2\cN^*\otimes\cK_3)=H^3(\cA,\cO_\cA({\bf k}_3-{\bf q}))$ to be 
non-trivial the vector ${\bf k}_3$ is required to have  precisely three entries less than or equal to $0$ and the others greater than or equal to $2$. 
Due to Eq.~\eqref{kzero}, these non-positive entries must arise in different components of the three vectors. 
Without restricting generality, we can, therefore, 
assume that $k_1^1\leq -2$, with all other components of ${\bf k}_1$ being greater than or equal to $0$,  $k_2^2\leq -2$ 
with all other components of ${\bf k}_2$ greater than or equal to $0$,  $k_3^3\leq 0$,  $k_3^4\leq 0$, $k_3^5\leq 0$, while $k_3^1\geq 2$, $k_3^2\geq 2$.
Using these conventions, we can specialise Eq.~\eqref{formsaff} to find the following explicit expressions for the relevant ambient space forms:
\be 
 \hat{\nu}_1 = \kappa^{k^1_1}_1 P_{(\mathbf{k_1})} d \overline{z}_1\; ,\qquad
 \hat{\nu}_2 = \kappa^{k^2_2}_2 Q_{(\mathbf{k_2})} d \overline{z}_2\;, \qquad
 \hat{\rho}_3 = \kappa^{k^3_3-2}_3 \kappa^{k^4_3-2}_4 \kappa^{k^5_3-2}_5 R_{(\mathbf{k_3}-\mathbf{q})} d \overline{z}_3 \wedge d \overline{z}_4 \wedge d \overline{z}_5 .
\ee
Inserting these forms into Eq.~\ref{yuk113} leads to
\be 
\l(\nu_1,\nu_2,\nu_3) =
\frac{1}{(2 \pi )^2}\int_{{\mathbb{C}^5}} d^5 z \ d^5\overline{z} \ \kappa^{k^1_1}_1 \kappa^{k^2_2}_2 \kappa^{k^3_3-2}_3 \kappa^{k^4_3-2}_4 \kappa^{k^5_3-2}_5 \ P_{(\mathbf{k_1})} Q_{(\mathbf{k_2})} R_{(\mathbf{k_3}-\mathbf{q})}.
 \label{5.10}
 \ee
By inserting expressions for the polynomials, this integral splits up into products of integrals over $\mathbb{P}^1$ and can be worked explicitly. Alternatively, we can proceed by noticing that the integrand $\hat{\nu}_1 \wedge \hat{\nu}_2 \wedge \hat{\rho}_3$ represents a cohomology class in $H^5(\cA,\cO_\cA(-{\bf q}))$, which is one-dimensional. Its harmonic representative has the form
\be
 c\,\mu(P,Q,R)\, \kappa_1^{-2} \kappa_2^{-2} \kappa_3^{-2} \kappa_4^{-2} \kappa_5^{-2} d^5 \overline{z}
\ee 
where 
\be
 \mu(P,Q,R)=\tilde{P}\tilde{Q}\tilde{R}\label{mu113}
\ee
must be a number, since $h^5(\cA,\cO_\cA(-{\bf q}))=1$. This number is obtained from polynomial multiplication as discussed in the previous sub-section and $c$ is a constant given by
\be 
c= c_{k_1^1,-k_1^1-2} \ c_{k_2^2,-k_2^2-2} \ c_{k_3^3-2,-k_3^3} \ c_{k_3^4-2,-k_3^4} \ c_{k_3^5-2,-k_3^5}\; .
\ee
Together with the basic identity 
\be
 \int_{\mathbb{C}} \dfrac{1}{\kappa^2} d z \wedge d \overline{z} = 2 \pi i\; ,
\ee 
this leads to the final expression 
\be 
\lambda(\nu_1,\nu_2,\nu_3) = 8 i \pi^3 c \ \mu(P,Q,R)\; , \qquad \mu(P,Q,R)=\tilde{P}\tilde{Q}\tilde{R}\; .\qquad  \label{alg113}
\ee
This equation represents our final result for the Yukawa couplings in this case and it allows for an ``algebraic'' calculation by multiplying together the polynomials $\tilde{P}$, $\tilde{Q}$ and $\tilde{R}$. Note that, given the rules for converting coordinates into partial derivatives in these polynomials, as discussed in the last sub-section, this must always result in a number, that is, the partial derivatives remove all remaining coordinates. 

\subsection{Yukawa couplings of type $(1,2,2)$}
The other simple case which avoids the vanishing theorem~\eqref{vanth2} arises 
if one of the forms, say $\nu_1$, is of type 1 while $\nu_2$ and  $\nu_3$ are of type 2. 
This case can be dealt with in complete analogy with the $(1,1,3)$ case in the previous sub-section.  
The relevant formula for the Yukawa couplings in this case is Eq.~\eqref{yuk122} which only involves 
the ambient space forms $\hat{\nu}_1\in H^1(\cA,\cK_1)$, $\hat{\o}_2\in H^2(\cA,\cN^*\otimes\cK_2)$ and $\hat{\o}_3\in H^2(\cA,\cN^*\otimes\cK_3)$.

In order to construct these forms it is again useful to fix our conventions. Since we require that $H^1(\cA,\cK_1)$ be non-trivial we need 
precisely one component in ${\bf k}_1$ less than or equal to $-2$ (and all others non-negative) 
and we choose $k_1^1\leq -2$. The two $(0,2)$ forms $\hat{\omega}_2$, $\hat{\omega}_3$ need to originate from 
different line bundles in the rank two bundle $\cN^*\otimes\cK_3$ (or else the Yukawa coupling would vanish) 
so we assume that $\hat{\omega}_2\in H^2(\cA,\cO_\cA(-{\bf q}_1)\otimes \cK_3)$ and $\hat{\omega}_3\in H^2(\cA,\cO_\cA(-{\bf q}_2)\otimes \cK_3)$. 
Hence we need precisely two entries in ${\bf k}_3-{\bf q}_1$ and  in ${\bf k}_3-{\bf q}_2$ to be less than or equal to $-2$ 
(with all other entries non-negative). Due to Eq.~\eqref{kzero} all 
negative entries have to arise in different components. 
Hence, we can choose  $k_2^2 - q_1^2 \leq -2$, $k_2^3 - q_1^3 \leq -2$, $k_3^4 - q_2^4 \leq -2$ and $k_3^5 - q_2^5 \leq -2$, with all the other entries 
non-negative. Applying these conventions to Eq.~\eqref{formsaff} results in
\bea 
 \hat{\nu}_1 &=& \kappa^{k^1_1}_1 P_{(\mathbf{k_1})} d \overline{z}_1\\
 \hat{\omega}_2 &=& \kappa^{k^2_2 - p^2_1}_2 \kappa^{k^3_2 - p^3_1}_3 Q_{(\mathbf{k}_2-\mathbf{p}_1)} d \overline{z}_2 \wedge d \overline{z}_3\\
 \hat{\omega}_3 &=& \kappa^{k^4_3-p^4_2}_4 \kappa^{k^5_3-p^5_2}_5  R_{(\mathbf{k_3}-\mathbf{p}_2)} d \overline{z}_4 \wedge d \overline{z}_5 \; .
\eea
Inserting these forms into Eq.~\eqref{yuk122}, the integral can be carried out as in the previous subsection and results in the same formula
\be 
\lambda(\nu_1,\nu_2,\nu_3) = 8 i \pi^3 c \ \mu(P,Q,R)\; , \qquad \mu(P,Q,R)=\tilde{P}\tilde{Q}\tilde{R}\; ,\qquad  \label{alg122}
\ee
but with the constant $c$ now given by
\be 
c= c_{k_1^1,-k_1^1-2} \ c_{k_2^2 - p^2_1 ,-k_2^2+p^2_1-2} \ c_{k_2^3 - p^3_1 ,-k_2^3+p^3_1-2} \ c_{k_3^4 - p^4_2 ,-k_3^4+p^4_2-2} \ c_{k_3^5 - p^5_2 ,-k_3^5+p^5_2-2}\; .
\ee
 
\subsection{An example with vanishing Yukawa couplings}
We consider a rank five line bundle sum on the CICY \eqref{conf7487} specified by the following line bundles:
\begin{align} 
L_1&=\mathcal{O}_X(1,0,-2,0,1), \,\,\,\,\,\,\,\,\,\,\,\, L_2=\mathcal{O}_X(1,-2,0,1,0), \,\,\,\,\,\,\,\,\,\,\,\, L_3=\mathcal{O}_X(0,1,0,0,-1)\notag \\
L_4&=\mathcal{O}_X(0,0,1,-1,0), \,\,\,\,\,\,\,\,\,\,\,\,  L_5=\mathcal{O}_X(-2,1,1,0,0)\; .
\end{align}
This model leads to a four-dimensional theory with gauge group $SU(5) \times S(U(1)^5)$. 
The non-vanishing cohomologies of these line bundles and their tensor products are
\be
\begin{array}{lllllll}
h^\bullet(X,L_1)&=&(0, 4, 0, 0) && h^\bullet(X,L_2)&=&(0, 4, 0, 0) \\
h^\bullet(X,L_5)&=&(0, 4, 0, 0)&& h^\bullet(X,L_1 \otimes L_2)&=&(0, 4, 0, 0) \\
h^\bullet(X,L_1 \otimes L_3)&=&(0,4,0,0) &&  h^\bullet(X,L_2 \otimes L_4)&=&(0,4,0,0)   \\
h^\bullet(X,L_3 \otimes L_4)&=&(0,1,1,0)  && h^\bullet(X,L_1 \otimes L_2^*)&=&(0,4,4,0)\\
h^\bullet(X,L_1 \otimes L_4^*)&=&(0, 16, 0, 0) && h^\bullet(X,L_2 \otimes L_3^*)&=&(0, 16, 0, 0)\\
h^\bullet(X,L_3 \otimes L_4^*)&=&(0, 1, 1, 0) && h^\bullet(X,L_5 \otimes L_3^*)&=&(0, 4, 0, 0)  \\
h^\bullet(X,L_5 \otimes L_4^*)&=&(0, 4, 0, 0)\; .
\end{array}
\ee
These results imply the following upstairs spectrum
\begin{align}
& 4 \ {\bf 10}_1,\,\, 4 \ {\bf 10}_2,\,\, 4 \ {\bf 10}_5, \notag \\ &  4 \ \overline{\bf 5}_{1,2}, \,\, 4 \ \overline{\bf 5}_{1,3}, \,\, 4 \ \overline{\bf 5}_{2,4}, \,\, \overline{\bf 5}^H_{3,4}, \,\,  {\bf 5}^{\overline{H}}_{3,4}, \notag \\ &  4 \ {\bf 1}_{1,2}, \,\,  4 \ {\bf 1}_{2,1}, \,\, 4 \ {\bf 1}_{1,3}, \,\, 12 \ {\bf 1}_{1,4}, \,\,  12 \ {\bf 1}_{2,3}, \,\,  4 \ {\bf 1}_{2,4}, \,\, {\bf 1}_{3,4}, \,\, {\bf 1}_{4,3}, \,\,  4 \ {\bf 1}_{5,3}, \,\,  4 \ {\bf 1}_{5,4}\; .
\end{align}
Here, the bold-face numbers denote $SU(5)$ representations and the subscripts indicate under which of the five $U(1)$ symmetries a multiplet is charged. This spectrum consists of $12$ families in $\bar{\bf 5}\oplus {\bf 10}$, one ${\bf 5}$--$\bar{\bf 5}$ pair of Higgs multiplets and a number of $SU(5)$ singlets. Upon dividing by the freely-acting symmetry $\Gamma = \mathbb{Z}_2 \times \mathbb{Z}_2$ in Eq.~\eqref{Z2Z2}, one obtains the standard model spectrum with three families. It is important to remember, however, that only couplings which respect the $S(U(1)^5)$ symmetry are allowed in the four-dimensional theory. One such allowed coupling is described by the following superpotential term:
\be
W =  \l_{I J K} \overline{\mathbf{5}}^{(I)}_{1,3} \overline{\mathbf{5}}^{(J)}_{2,4} \mathbf{10}^{(K)}_5\; .  \label{Wex1}
\ee
In order to compute this coupling, we write down the relevant line bundles and bundle-valued forms which are given by
\be
\begin{array}{ll}
4 \ \overline{\mathbf{5}}_{1,3} \rightarrow K_1=L_1 \otimes L_3 = \mathcal{O}_X(1, 1, -2, 0, 0), & \hat{\nu}_1= \sigma_3^{-2} \tilde{P}_{(1, 1, -2, 0, 0)} d\bar{\mu}_3 \in H^1(\mathcal{A},\mathcal{K}_1) \\
4 \ \overline{\mathbf{5}}_{2,4} \rightarrow K_2 = L_2 \otimes L_4 = \mathcal{O}_X(1, -2, 1, 0, 0), & \hat{\nu}_2= \sigma_2^{-2}  \tilde{Q}_{(1, -2, 1, 0, 0)} d\bar{\mu}_2\in  H^1(\mathcal{A},\mathcal{K}_2) \\
4 \ \mathbf{10}_5 \rightarrow K_3 = L_5  = \mathcal{O}_X(-2,1,1,0,0) , &  \hat{\nu}_3=\sigma_1^{-2}  \tilde{R}_{(-2,1,1,0,0)} d\bar{\mu}_1 \in  H^1(\mathcal{A},\mathcal{K}_3)\; ,
\end{array}\label{nuex1}
\ee
with explicit polynomials
\begin{align}
\tilde{P} & =  p_0 x_0 y_0+p_1 x_0 y_1 + p_2 x_1 y_0 + p_3 x_1 y_1 \nonumber  \\
\tilde{Q} & = q_0 x_0 u_0+q_1 x_0 u_1 + q_2 x_1 u_0 + q_3 x_1 u_1  \\
\tilde{R} & = r_0 y_0 u_0+r_1 y_0 u_1 + r_2 y_1 u_0 + r_3 y_1 u_1\; . \nonumber
\end{align}
Evidently, from Eq.~\eqref{nuex1}, all three forms $\nu_i$ are of type $\tau_i=1$ and, hence, the Yukawa couplings $\l_{IJK}$ in Eq.~\eqref{Wex1} are all zero as a consequence of the vanishing theorem~\eqref{vanth2}.

\subsection{An example with Yukawa couplings of type $(1,1,3)$}
\label{section311}
A line bundle model on the CICY  \eqref{conf7487} which realises Yukawa couplings of type $(\tau_1,\tau_2,\tau_3)=(1,1,3)$ is defined by the five line bundles
\begin{align} 
L_1&=\mathcal{O}_X(1,-2,0,0,1), \,\,\,\,\,\,\,\,\,\,\,\, L_2=\mathcal{O}_X(0,1,0,1,-2), \,\,\,\,\,\,\,\,\,\,\,\, L_3=\mathcal{O}_X(0,0,1,-2,1) \notag \\
L_4&=\mathcal{O}_X(0,0,-1,0,1), \,\,\,\,\,\,\,\,\,\,\,\,  L_5=\mathcal{O}_X(-1,1,0,1,-1)\; .
\end{align}
As before, the four-dimensional gauge group is $SU(5) \times S(U(1)^5)$ and the non-trivial cohomologies of the above line bundles and their tensor product
\begin{eqnarray}
\begin{array}{lllllll}
h^\bullet(X,L_1)&=&(0, 4, 0, 0) && h^\bullet(X,L_2)&=&(0, 4, 0, 0) \\
h^\bullet(X,L_3)&=&(0, 4, 0, 0)  && h^\bullet(X,L_1 \otimes L_3)&=&(0, 4, 0, 0) \\
h^\bullet(X,L_2 \otimes L_3)&=&(0,1,1,0) &&  h^\bullet(X,L_2 \otimes L_4)&=&(0,1,1,0)   \\
h^\bullet(X,L_2 \otimes L_5)&=&(0,8,0,0)  && h^\bullet(X,L_3 \otimes L_4)&=&(0,3,3,0)  \\
h^\bullet(X,L_1 \otimes L_4^*)&=&(0, 4, 0, 0) && h^\bullet(X,L_1 \otimes L_5^*)&=&(0, 8, 0, 0) \\
h^\bullet(X,L_2 \otimes L_3^*)&=&(0, 9, 9, 0) && h^\bullet(X,L_2 \otimes L_4^*)&=&(0, 16, 0, 0) \\
h^\bullet(X,L_3 \otimes L_4^*)&=&(0, 3, 3, 0) && h^\bullet(X,L_3 \otimes L_5^*)&=&(0, 12, 0, 0) \\
h^\bullet(X,L_5 \otimes L_4^*)&=&(0, 4, 0, 0) 
\end{array}
\end{eqnarray}
lead to the following spectrum:
\begin{align}
& 4 \ {\bf 10}_1,\,\, 4 \ {\bf 10}_2,\,\, 4 \ {\bf 10}_3, \notag \\ &  4 \ \overline{\bf 5}_{1,3},  \,\, \ \overline{\bf 5}^H_{2,3}, \,\,  \ {\bf 5}^{\overline{H}}_{2,3}, \,\, \ \overline{\bf 5}^H_{2,4}, \,\,  \ {\bf 5}^{\overline{H}}_{2,4}, \,\, 8 \ \overline{\bf 5}_{2,5},  \,\, 3 \ \overline{\bf 5}^H_{3,4}, \,\,  3 \ {\bf 5}^{\overline{H}}_{3,4},  \notag \\ &  4 \ {\bf 1}_{1,4}, \,\, 8 \ {\bf 1}_{1,5}, \,\, 9 \ {\bf 1}_{2,3}, \,\,  9 \ {\bf 1}_{3,2}, \,\,  16 \ {\bf 1}_{2,4}, \,\,  3 \ {\bf 1}_{3,4}, \,\,  3 \ {\bf 1}_{4,3}, \,\, 12 \ {\bf 1}_{3,5}, \,\,  4 \ {\bf 1}_{5,4}\; .
\end{align}
This spectrum contains $12$ families $\bar{\bf 5}\oplus{\bf 10}$, five ${\bf 5}$--$\bar{\bf 5}$ Higgs pairs and $SU(5)$-singlet multiplets and gives rise to a three-family standard model after a suitable quotient with the symmetry~\eqref{Z2Z2}. We are interested in the superpotential terms
\begin{align}
W & =  \l_{I J K}  \overline{\mathbf{5}}^{H,(I)}_{3,4}\mathbf{10}^{(J)}_1\overline{\mathbf{5}}^{(K)}_{2,5} \; .  \label{Wex2}
\end{align}
which are allowed by all gauge symmetries of the model. The relevant harmonic forms are given by
\begin{eqnarray}
\begin{array}{ll}
3 \ \overline{\mathbf{5}}^H_{3,4} \rightarrow K_1 = L_3 \otimes L_4 = \mathcal{O}_X(0, 0, 0, -2, 2), & \hat{\nu}_1=  \sigma_4^{-2}  \tilde{P}_{(0, 0, 0, -2, 2)} d\bar{\mu}_4  \\
4 \ \mathbf{10}_1 \rightarrow K_2 = L_1  = \mathcal{O}_X(1,-2,0,0,1), &  \hat{\nu}_2=\sigma_2^{-2}  \tilde{Q}_{(1,-2,0,0,1)} d\bar{\mu}_2 \\
8 \ \overline{\mathbf{5}}_{2,5} \rightarrow K_3=L_2 \otimes L_5 = \mathcal{O}_X(-1, 2, 0, 2, -3), & \hat{\rho}_3= \sigma_1^{-3} \sigma_3^{-2} \sigma_5^{-5} \tilde{R}_{(-3,0,-2,0,-5)} d\bar{\mu}_1 \wedge d\bar{\mu}_3 \wedge d\bar{\mu}_5 \\
\end{array}
\end{eqnarray}
where $\hat{\nu}_1 \in H^1(\mathcal{A},\mathcal{K}_1)$,  $\hat{\nu}_2 \in H^1(\mathcal{A},\mathcal{K}_2)$ and $\hat{\rho}_3 \in H^3(\mathcal{A},\wedge^2 \mathcal{N}^* \otimes \mathcal{K}_3)$. The polynomials $\tilde{P}$, $\tilde{Q}$, $\tilde{R}$ can be explicitly written as
\begin{eqnarray}
\tilde{P} & =& p_0 w_0^2 + p_1 w_0 w_1 + p_2 w_1^2  \notag  \\
\tilde{Q} & = & q_0 x_0 w_0 + q_1 x_1 w_0 + q_2 x_0 w_1 + q_3 x_1 w_1\\
\tilde{R} & = & r_0 \overline{x}_0 \overline{w}^3_0 + r_1 \overline{x}_0 \overline{w}^2_0 \overline{w}_1 + r_2  \overline{x}_0 \overline{w}_0 \overline{w}^2_1 + r_3 \overline{x}_0 \overline{w}^3_1 +\notag \\
 && r_4 \overline{x}_1 \overline{w}^3_0 + r_5 \overline{x}_1 \overline{w}^2_0 \overline{w}_1 + r_6  \overline{x}_1 \overline{w}_0 \overline{w}^2_1 + r_7 \overline{x}_1 \overline{w}^3_1\; . \notag 
\end{eqnarray}
Note that the coefficients $p_I$, $q_J$ and $r_K$ in these polynomials parametrise the various families. Using these polynomials we can compute the upstairs Yukawa couplings from Eq.~\eqref{alg113} which leads to
\begin{eqnarray}
\mu(\tilde{P},\tilde{Q},\tilde{R})& =& 6 p_0 q_0 r_0 + 2 p_1 q_0 r_1  + 2 p_2 q_0 r_2  + 6 p_0 q_1 r_4 + 2 p_1 q_1 r_5 + 2 p_2 q_1 r_6 + \notag \\ &&  2 p_0 q_2 r_1 + 2 p_1 q_2 r_2 + 6 p_2 q_2 r_3 + 2 p_0 q_3 r_5 +  2 p_1 q_3 r_6 + 6 p_2 q_3 r_7\; .
\end{eqnarray}
The Yukawa couplings $\l_{IJK}$ in Eq.~\eqref{Wex2} can be easily obtained from this expression by choosing a basis for the coefficients, for example by setting one each of the coefficients $p_I$, $q_J$, $r_K$ to one and the others to zero. It is however more interesting to see what happens in the downstairs theory, obtained from the present $SU(5)$ GUT theory by a quotient with the $\G=\mathbb{Z}_2\times \mathbb{Z}_2$ symmetry~\eqref{Z2Z2}.  The GUT multiplets branch as $\mathbf{10}\rightarrow (Q,u,e)$, $\overline{\mathbf{5}}\rightarrow (d,L)$, $\overline{\mathbf{5}}^H\rightarrow (T,H)$ into standard model multiplets, where $T$ is the Higgs triplet which has to be projected out. 
On the quotient manifold $\tilde{X}$ we introduce a Wilson line in the standard hypercharge direction in order to break $SU(5)$ to the standard model group. Such a Wilson line can be described by two $\G$-representations $\chi_2$, $\chi_3$ which we choose as $\chi_2=(1,1)$ and $\chi_3=(0,0)$. This induces the multiplet charges
\be
  \chi_d = \chi_3^* = (0,0)\; ,\qquad \chi_H = \chi_2^* = (1,1)\; ,\qquad \chi_Q  = \chi_2 \otimes \chi_3 = (1,1)\; . \label{Wilsonex2}
\ee
\noindent In order to determine the polynomials corresponding to the downstairs spectrum, one has to keep in mind that every differential $d \overline{\mu}_i$ has charge $(1,1)$ under $\G$. Moreover, for the $(0,3)$-form $\hat{\rho}_3$, there is an additional $(1,1)$ charge flip due to the fact that the bundle $ \wedge^2 \mathcal{N}^*\otimes \cK_3 $ transforms non-trivially under $\G$ from Eq.~\eqref{Z2Z2pol}. Matching these charges up with the Wilson line charges~\eqref{Wilsonex2} the representatives of the downstairs spectrum become
\begin{eqnarray}
H_{3,4}&:& \tilde{P} \in{\rm Span}(w_0^2 + w_1^2) \notag \\
Q_1&:&\tilde{Q} \in{\rm Span}(x_0 w_0 + x_1 w_1) \\
d_{2,5}&:&\tilde{R} \in{\rm Span}( \overline{x}_0 \overline{w}_0 \overline{w}_1^2 + \overline{x}_1 \overline{w}_1 \overline{w}_0^2, \overline{x}_0 \overline{w}_0^3 + \overline{x}_1 \overline{w}_1^3)  \notag 
\end{eqnarray}
Using Eq.~\eqref{alg113} the Yukawa couplings in $\l_KH_{3,4}Q_1d_{2,4}^{(K)}$ become proportional to
\begin{eqnarray}
\mu(H_{3,4}, Q_1, d_{2,5}^{(K)}) =  \dfrac{1}{4} \left( \partial^2_{\overline{w}_0}+ \partial^2_{\overline{w}_1}\right) \left( \partial_{\overline{x}_0}\partial_{\overline{w}_0}+ \partial_{\overline{x}_1}\partial_{\overline{w}_1}\right) \begin{pmatrix} \overline{x}_0 \overline{w}_0 \overline{w}_1^2 + \overline{x}_1 \overline{w}_1 \overline{w}_0^2 \\ \overline{x}_0 \overline{w}_0^3 + \overline{x}_1 \overline{w}_1^3 \end{pmatrix} = \begin{pmatrix} 1 \\ 3 \end{pmatrix},
\end{eqnarray}
where we have converted the homogeneous coordinates into derivatives and introduced an additional  factor $1/4$, to account for the fact that the integral is carried out on the quotient manifold. The numerical coefficient $c$ in Eq.~\eqref{alg113} is given by
\begin{eqnarray}
c = c_{(-2,0)} c_{(-2,0)} c_{(-5,3)} c_{(-4,2)}  c_{(-7,5)} = 1\cdot 1\cdot\dfrac{1}{4!}\cdot\dfrac{1}{3!}\cdot\dfrac{1}{6!}
\end{eqnarray}

\subsection{An example with Yukawa couplings of type $(1,2,2)$}
This example on the CICY~\eqref{conf7487} is defined by the five line bundles
\begin{align} 
L_1&=\mathcal{O}_X(1,-2,0,0,1), \,\,\,\,\,\,\,\,\,\,\,\, L_2=\mathcal{O}_X(0,1,-2,0,1), \,\,\,\,\,\,\,\,\,\,\,\, L_3=\mathcal{O}_X(0,0,1,1,-2) \notag \\
L_4&=\mathcal{O}_X(0,0,1,-1,0), \,\,\,\,\,\,\,\,\,\,\,\,  L_5=\mathcal{O}_X(-1,1,0,0,0)\; .
\end{align}
The non-vanishing cohomologies of these line bundles and their tensor products
\be
\begin{array}{lllllll}
h^\bullet(X,L_1)&=&(0, 4, 0, 0) && h^\bullet(X,L_2)&=&(0, 4, 0, 0) \\
h^\bullet(X,L_3)&=&(0, 4, 0, 0)  && h^\bullet(X,L_1 \otimes L_3)&=&(0, 4, 0, 0) \\
h^\bullet(X,L_1 \otimes L_4)&=&(0,4,0,0) &&  h^\bullet(X,L_2 \otimes L_3)&=&(0,1,1,0)   \\
h^\bullet(X,L_2 \otimes L_4)&=&(0,1,1,0)  && h^\bullet(X,L_3 \otimes L_4)&=&(0,3,3,0)  \\
h^\bullet(X,L_3 \otimes L_5)&=&(0, 4, 0, 0) && h^\bullet(X,L_1 \otimes L_2^*)&=&(0, 12, 0, 0) \\
h^\bullet(X,L_3 \otimes L_1^*)&=&(0, 12, 0, 0) && h^\bullet(X,L_1 \otimes L_4^*)&=&(0, 4, 0, 0) \\
h^\bullet(X,L_1 \otimes L_5^*)&=&(0, 12, 0, 0) && h^\bullet(X,L_2 \otimes L_3^*)&=&(0, 9, 9, 0) \\
h^\bullet(X,L_2 \otimes L_4^*)&=&(0, 16, 0, 0) && h^\bullet(X,L_2 \otimes L_5^*)&=&(0, 4, 0, 0) \\
h^\bullet(X,L_3 \otimes L_4^*)&=&(0, 3, 3, 0) && h^\bullet(X,L_3 \otimes L_5^*)&=&(0, 4, 0, 0)
\end{array}
\ee
lead to the upstairs spectrum
\begin{align}
& 4 \ {\bf 10}_1,\,\, 4 \ {\bf 10}_2,\,\, 4 \ {\bf 10}_3, \notag \\ &  4 \ \overline{\bf 5}_{1,3}, \,\, 4 \ \overline{\bf 5}_{1,4}, \,\, \ \overline{\bf 5}^H_{2,3}, \,\,  \ {\bf 5}^{\overline{H}}_{2,3}, \,\, \ \overline{\bf 5}^H_{2,4}, \,\,  \ {\bf 5}^{\overline{H}}_{2,4},  \,\, 3 \ \overline{\bf 5}^H_{3,4}, \,\,  3 \ {\bf 5}^{\overline{H}}_{3,4}, \,\, 4 \ \overline{\bf 5}_{3,5}, \notag \\ &  12 \ {\bf 1}_{1,2}, \,\, 12 \ {\bf 1}_{3,1}, \,\, 4 \ {\bf 1}_{1,4}, \,\,  12 \ {\bf 1}_{1,5}, \,\,  9 \ {\bf 1}_{2,3}, \,\,  9 \ {\bf 1}_{3,2}, \,\,  16 \ {\bf 1}_{2,4}, \,\, 4 \ {\bf 1}_{2,5}, \,\,  3 \ {\bf 1}_{3,4}, \,\,  3 \ {\bf 1}_{4,3}, \,\,  4 \ {\bf 1}_{3,5}\; .
\end{align}
As before, this spectrum with 12 families in $\bar{\bf 5}\oplus{\bf 10}$, five ${\bf 5}$--$\bar{\bf 5}$ Higgs pairs and $SU(5)$-singlets leads to a three-family standard model after the quotient by $\G=\mathbb{Z}_2\times\mathbb{Z}_2$. We are interested in the following superpotential term:
\be
W  =  \l_{I J K} \mathbf{10}^{(I)}_2  \overline{\mathbf{5}}^{(J)}_{1,4} \overline{\mathbf{5}}^{(K)}_{3,5}
\ee
The associated harmonic forms 
\be
\begin{array}{lll}
4 \ \mathbf{10}_2 \rightarrow K_1 = L_2  = \mathcal{O}_X(0,1,-2,0,1), && \hat{\nu}_1=\sigma_3^{-2}  P_{(0,1,-2,0,1)} d\bar{\mu}_3 \\
4 \ \overline{\mathbf{5}}_{1,4} \rightarrow K_2=L_1 \otimes L_4 = \mathcal{O}_X(1,-2,1,-1,1), && \hat{\omega}_2= \sigma_2^{-3} \sigma_4^{-2} Q_{(1, -3, 0, -2, 0)} d\bar{\mu}_2 \wedge d\bar{\mu}_4  \\
4 \ \overline{\mathbf{5}}_{3,5} \rightarrow K_3 = L_3 \otimes L_5 = \mathcal{O}_X(-1,1,1,1,-2), && \hat{\omega}_3= \sigma_1^{-3} \sigma_5^{-3}  R_{(-3, 0, 0, 0, -3)} d\bar{\mu}_1 \wedge d\bar{\mu}_5\; ,  \\
\end{array}
\ee
where $\hat{\nu}_1 \in H^1(\mathcal{A},\mathcal{K}_1$,   $\hat{\omega}_2 \in H^2(\mathcal{A},\mathcal{N}^* \otimes \mathcal{K}_2)$ and $\hat{\omega}_3 \in H^2(\mathcal{A},\mathcal{N}^* \otimes \mathcal{K}_3)$, contain the polynomials
\begin{align}
\tilde{P} & = p_0 y_0 w_0 + p_1 y_1 w_0 + p_2 y_0 w_1 + p_3 y_1 w_1\\
\tilde{Q} & =  q_0 x_0 \overline{y}_0 + q_1 x_1 \overline{y}_0 + q_2 x_0 \overline{y}_1  + q_3 x_1 \overline{y}_1  \notag  \\
\tilde{R} & = r_0 \overline{x}_0 \overline{w}_0 + r_1 \overline{x}_1 \overline{w}_0  + r_2 \overline{x}_0 \overline{w}_1  + r_3 \overline{x}_1 \overline{w}_1\; .   \notag  \\
\end{align}
From Eq.~\eqref{alg122} this leads to upstairs Yukawa couplings
\be
\mu(\tilde{P},\tilde{Q},\tilde{R}) = p_0 q_0 r_0 +  p_1 q_0 r_1 +  p_2 q_3 r_2 + p_3 q_3 r_3\; .
\ee
Under the standard model group, the relevant part of the upstairs spectrum branches as $\mathbf{10}_2\rightarrow (Q,u,e)_2$, $\overline{\mathbf{5}}_{1,4}\rightarrow (d,L)_{1,4}$, $\overline{\mathbf{5}}_{3,5}\rightarrow (T,H)_{3,5}$. We choose the same Wilson line, $\chi_2=(1,1)$ and $\chi_3=(0,0)$, as in Section~\ref{section311}, which then leads to the same multiplet charges as in Eq.~\eqref{Wilsonex2}. Once again, we have to keep in mind that the differentials $d \overline{\mu}_i$ carry charge $(1,1)$ under $\G$. Moreover, we have to remember from Eq.~\eqref{Z2Z2pol} that forms which arise from $\cO_\cA(-{\bf q}_2)\otimes\cK_i$ transform with an additional overall $\G$-charge $(1,1)$, while forms from $\cO_\cA(-{\bf q}_1)\otimes\cK_i$ do not. With these rules, the polynomials corresponding to the downstairs spectrum turn out to be
\begin{eqnarray}
Q_2&:& \tilde{P} \in{\rm Span}(y_0 w_0 + y_1 w_1)\nonumber\\  
d_{1,4}&:& \tilde{Q}\in{\rm Span}( x_0 \overline{y}_0 + x_1 \overline{y}_1)  \\
H_{3,5}&:&\tilde{R} \in{\rm Span}(\overline{x}_0 \overline{w}_0 + \overline{x}_1 \overline{w}_1) \nonumber
\end{eqnarray}
Then, from Eq.~\eqref{alg122}, the downstairs Yukawa coupling for $H_{3,4}\, Q_1\, d_{2,5}$ is proportional to
\be
\mu(H_{3,4}, Q_1, d_{2,5}) = \dfrac{1}{4}  \left( \partial_{\overline{x}_0}\partial_{\overline{w}_0}+ \partial_{\overline{x}_1} \partial_{\overline{w}_1}\right) \left( x_0 \partial_{\overline{y}_0}+ x_1 \partial_{\overline{y}_1}\right)  \left( y_0 w_0 + y_1 w_1 \right) = \dfrac{1}{2} 
\ee
with the constant $c$ given by
\be
c = c_{(-2,0)} c_{(-4,2)} c_{(-3,1)} c_{(-4,2)} c_{(-5,3)} = 1 \cdot \dfrac{1}{3!} \cdot \dfrac{1}{2!} \cdot \dfrac{1}{3!} \cdot \dfrac{1}{4!}\; .
\ee

\section{Conclusion}\label{conclusion}
In Ref.~\cite{Blesneag:2015pvz} methods to calculate the holomorphic Yukawa couplings have been developed for line bundle models on certain special Calabi-Yau manifolds, with a focus on the tetra-quadric Calabi-Yau manifolds defined in the ambient space $\mathbb{P}^1\times \mathbb{P}^1\times\mathbb{P}^1\times\mathbb{P}^1$. The present paper generalises these methods to all CICY manifolds, and, hence, demonstrates that they are applicable to large classes of Calabi-Yau manifolds. 

Our methods rely on the presence of an ambient space $\cA$, presently a product of projective spaces although generalisations are likely possible, into which the Calabi-Yau manifold $X$ is embedded at co-dimension $k$. Likewise, the three line bundles $K_i\rightarrow X$ associated to a Yukawa coupling should be restrictions of ambient space line bundles $\cK_i\rightarrow \cA$. We have shown that, in this situation, the three $K_i$-valued $(0,1)$--forms $\nu_i\in H^1(X,K_i)$ which enter the expression for the holomorphic Yukawa couplings can each be related to a chains $\hat{\nu}_{i,a}$ of $(0,a)$ ambient space forms, where $a=1,\ldots ,k+1$. Moreover, from Eq.~\eqref{yukfin}, the Yukawa couplings can be written in terms of these ambient space forms $\hat{\nu}_{i,a}$.

We say that a form $\nu_i$ is of type $\tau_i\in\{1,\ldots ,k+1\}$ if it originates from the ambient space $(0,\tau_i)$-form $\nu_{i,\tau_i}\in H^{\tau_i}(\cA,\wedge^{\tau_i-1}\cN^*\otimes\cK_i)$. This means that the associated chain of ambient space forms breaks down at $a=\tau_i$, that is, if $\hat{\nu}_{i,a}= 0$ for $a>\tau_i$. One of our main results is a vanishing theorem which states that the Yukawa coupling between three forms $\nu_i$ vanishes if their associated types satisfy $\tau_1+\tau_2+\tau_3<{\rm dim}(\cA)$. Especially for large ambient space dimension ${\rm dim}(\cA)$ this implies the vanishing of many Yukawa couplings since large types $\tau_i$ tend to be rare. This vanishing is not explained by one of the obvious four-dimensional symmetries and, therefore, appears to be topological in nature. 

The nature of this vanishing statement is somewhat puzzling in that it relates a physical property - the vanishing of Yukawa couplings - to conditions on unphysical quantities, essentially properties of the ambient space $\cA$ which is auxiliary and carries no physical relevance. We do not currently know if the vanishing statement is restricted to Calabi-Yau manifolds which can be embedded into an ambient space in this way or if it extends to all Calabi-Yau manifolds. In the latter case, there should be an ``intrinsic" formulation of this statement which only refers to properties of the Calabi-Yau manifold. 

We have illustrated our methods by computing certain holomorphic Yukawa couplings for three different line bundle standard models on a co-dimension two CICY. 
The most pressing issue is, of course, the calculation of the matter field K\"ahler potential and, hence, of the physical 
Yukawa couplings. We hope that the methods developed in this paper will help in this regard and work in this direction is currently underway.

\section*{Acknowledgements}
The work of E.I.B. was supported by the ARC Future Fellowship FT120100466 and in part by the ARC Discovery project DP140103925.
A.L. is partially supported by the EPSRC network grant EP/N007158/1 and by the STFC grant~ST/L000474/1.
E.I.B.~would like to thank physics department at Oxford University where some of this work was done for warm hospitality.
\newpage


\appendix
\section{The boundary integral}\label{appA}
In deriving the formula for the Yukawa couplings, we have performed a number of partial integrations, starting with Eq.~\eqref{yukstart}, in order to arrive at the final result~\eqref{yukfin}. The boundary terms from those partial integrations have not been taken into account. The purpose of this appendix is to justify this procedure by showing that the boundary terms do indeed vanish.

\subsection{The co-dimension one case}

We start with the ambient space
\be 
{\cal A}= {\mathbb P}^{n_1} \times {\mathbb P}^{n_2} \times \dots  {\mathbb P}^{n_m}\,, \qquad 
\sum_{i=1}^m n_i = 4
\label{A0}
\ee
and a Calabi-Yau hypersurface $X\subset \cA$ defined as the zero locus of a polynomial $p$ of multi-degree
$(n_1+1, \dots, n_m+1)$. The relevant integral for the Yukawa couplings, before the integration by parts, reads\footnote{In this Appendix we ignore various numeric pre-factors since they do not matter for our discussion.}
\be 
\l (\nu_1, \nu_2, \nu_3) \sim \int_{{\mathbb C}^4} d^4 z \wedge \hat{\nu}_1 \wedge \hat{\nu}_2 \wedge \hat{\nu}_3 
\wedge {\bar \pt} \Big(\frac{1}{p}\Big)\, , 
\label{A1}
\ee
where $z_1, \dots, z_4$ are affine coordinates on a patch ${\mathbb C}^4$ of $\cA$. Let us introduce the $(0,3)$ form
\be 
\hat{\a} =  \hat{\nu}_1 \wedge \hat{\nu}_2 \wedge \hat{\nu}_3  \in \Omega^3 ({\cal A}, {\cal O}_{{\cal A}})\; , 
\label{A2}
\ee
which takes values in the trivial bundle. Further, we define the form $\hat{\beta}$ by
\be 
\bar \pt \hat{\a} = p\hat{\b}\;.
\label{betadef1}
\ee
Note that $\hat{\b} \in H^4 ({\cal A}, {\cal O}_{{\cal A}} (-n_1-1, \dots, -n_m-1))\cong\mathbb{C}$ and, hence, that $\hat{\beta}$ is uniquely fixed up to an overall constant and an exact form, both of which are irrelevant for the present purposes. A harmonic representative for $\hat{\beta}$ can be written down following the rules in Appendix~\ref{appC} (for the case ${\cal A}= {\mathbb P}^{1} \times {\mathbb P}^{1} \times  {\mathbb P}^{1}\times  {\mathbb P}^{1}$ it was also constructed in~\cite{Blesneag:2015pvz}) and this leads to
\be 
\hat{\b} \sim \frac{d^4 {\bar z}}{ \kappa_1^{n_1+1} \dots \kappa_k^{n_k+1}}\;. 
\label{A5}
\ee
In order to understand the boundary integral we need to study the limit when the modulus of one of the coordinates, say $z_1$, goes to infinity. Let us assume that $z_1$ is an affine coordinate of the first projective factor ${\mathbb P}^{n_1}$. Then, for large $|z_1|$, we have
\be 
\hat{\b} \sim \frac{d^4 \bar z}{z_1^{n_1+1} {\bar z}_1^{n_1+1}} \,, \qquad p\b \sim \frac{d^4 \bar z}{ {\bar z}_1^{n_1+1}}\,. 
\label{A6}
\ee
Let us solve Eq.~\eqref{betadef1} for $\hat{\a}$ in this limit. The general solution for $\hat{\a}$ is given by 
$\hat{\a}=\hat{\a}_0+ \hat{\a}_1$, where $\hat{\a}_0$ is the general solution to the homogeneous equation $\bar\pt\hat{\a}=0$ and $\hat{\a}_1$ is a 
partial solution to the inhomogeneous equation~\eqref{betadef1}. For a four-dimensional ambient space of the form~\eqref{A0} we have $H^3 ({\cal A}, {\cal O}_{{\cal A}})=0$ and, hence, $\hat{\a}_0$ is exact and, therefore, irrelevant for the integral. From Eq.~\eqref{betadef1} we conclude that 
\be 
\hat{\a}=\hat{\a}_1 = \frac{1}{{\bar z}_1^{n_1}} \hat{\a}'\,, 
\label{A7}
\ee
where $\hat{\a}'$ is  a $(0, 3)$-form independent of $z_1, {\bar z}_1$ and $d {\bar z}_1$. Note that $\hat{\a} \to 0$ for large $|z_1|$. From Eq.~\eqref{A1} we find that the boundary term in the limit $|z_1| \to \infty$ behaves as
\be 
\int_{{\mathbb C}^3 \times \gamma_1} d^4z \wedge \frac{\hat{\a}}{p}\Big|_{|z_1| \to \infty}\,, 
\label{A8}
\ee
where $\gamma_1$ is the circle at infinity in the complex plane parameterised by $z_1$. This contour integral is zero since, generically, $p \sim z_1^{n_1+1}$, and $\hat{\a} \to 0$ for large $|z_1|$. 


\subsection{The co-dimension two case}

We will now repeat this discussion for a co-dimension two CICY with ambient space
\be 
{\cal A}= {\mathbb P}^{n_1} \times {\mathbb P}^{n_2} \times \dots  {\mathbb P}^{n_m}\,, \quad 
\sum_{i=1}^m n_i = 5\; .
\label{A9}
\ee
The CICY $X\subset \cA$ is defined as the common zero locus of a a pair of polynomials $ p = (p_1, p_2)$ with multi-degrees ${\bf q}_1=(q_1^1,\ldots ,q_1^m)$ and ${\bf q}_2=(q_2^1,\ldots ,q_2^m)$, satisfying the Calabi-Yau condition $q_1^i+q_2^i=n_i+1$, for all $i=1,\ldots ,m$. Introducing affine coordinates $(z_1,\ldots ,z_5)$ on a patch in $\cA$, the formula for the Yukawa coupling can be written as
\be 
\l \sim \int_{{\mathbb C}^5} d^5 z \wedge \hat{\a} \wedge \bar \pt \Big( \frac{1}{p_1}\Big) \wedge \bar \pt \Big( \frac{1}{p_2}\Big)\,, 
\label{A10}
\ee
where $\hat{\a}$ is given by Eq.~\eqref{A2}. Using the results from Section~\ref{codim2} we obtain 
\bea
&& 
\bar \pt \hat{\a} = p \hat{\b} = p_1 \hat{\b}^{1} + p_2 \hat{\b}^{2}\,, \nonumber \\
&&
\bar \pt \hat{\b}^{1}= - p_2 \hat{\eta}\,, \quad \bar \pt \hat{\b}^{2}=  p_1 \hat{\eta}\,, \nonumber \\
&&
\bar \pt \hat{\eta} =0\,.
\label{A13}
\eea
From Eqs.~\eqref{A2}, \eqref{A13} it follows that 
\be
\hat{\b}^{a} \in \Omega^4 ({\cal A}, {\cal O}_{{\cal A}}  (-{\bf d}^{a}))\,, \qquad
\hat{\eta} \in H^5 ({{\cal A}}, {\cal O}_{{\cal A}}  (-{\bf d}^{1}-{\bf d}^{2} )) = H^5 ({{\cal A}}, \Lambda^2 {\cal N}^*)\cong\mathbb{C}\,. 
\label{A14}
\ee
This means that the form $\hat{\eta}$ is unique up to a multiplicative coefficient and an exact form, both irrelevant in the present context. As in the previous subsection 
we can use the results from Appendix~\ref{appC} to write down the harmonic representative
\be 
\hat{\eta} \sim \frac{d^5 \bar z}{\kappa_1^{n_1+1} \dots \kappa_m^{n_m+1}}\,. 
\label{A15}
\ee
To compute the boundary integrals we need to study the behaviour in the limit when the modulus of one of the affine coordinates, say $z_1$, 
goes to infinity. Let us assume that $z_1$ is an affine coordinate of the first projective factor ${\mathbb P}^{n_1}$. In the large $|z_1|$ limit we obtain 
\be
\hat{\eta} \sim \frac{d^5 \bar z}{z_1^{n_1+1} {\bar z}_1^{n_1+1}}\;,\qquad
p_1 \hat{\eta}  \sim \frac{d^5 \bar z}{z_1^{q_2^{1}} {\bar z}_1^{n_1+1}}\;, \qquad 
p_2 \hat{\eta}  \sim \frac{d^5 \bar z}{z_1^{q_1^{2}} {\bar z}_1^{n_1+1}}\; . 
\label{A16}
\ee
Using Eq.~\eqref{A13} we can now obtain the behaviour of $\hat{\b}^{a}$ and $\hat{\a}$ in the limit of large $|z_1|$. Their general solution is given 
by 
\be 
\hat{\b}^{a}= \hat{\b}^{a}_0+ \hat{\b}^{a}_1 \,, \quad \hat{\a}= \hat{\a}_0+ \hat{\a}_1\,, 
\label{A17}
\ee
where $\hat{\b}^{a}_0$, $\hat{\a}_0$ are the general solutions to the corresponding homogeneous equations and $\hat{\b}^{a}_1$, $\hat{\a}_1$
are partial solutions to the inhomogeneous equations. For a 5-dimensional ambient space of the form~\eqref{A9} 
we have $H^{3}({\cal A}, {\cal O}_{{\cal A}})=0$ and $H^{4}({\cal A}, {\cal O}_{{\cal A}} (-{\bf d}^{a}))=0$ so that $\hat{\a}_0$ and $\hat{\b}^{a}_0$ are both exact and can be discarded. Solving for $\hat{\b}^{a}_1$ and $\hat{\a}_1$ yields
\be
\hat{\b}^{1}= \hat{\b}^{1}_1 \sim \frac{d {\bar  z}_2 \wedge \dots \wedge  d {\bar  z}_5}{z_1^{q_1^{1}}  {\bar z}_1^{n_1}}\;, \qquad 
\hat{\b}^{2}= \hat{\b}^{2}_1 \sim \frac{d {\bar  z}_2 \wedge \dots \wedge  d {\bar  z}_5}{z_1^{q_2^{1}}  {\bar z}_1^{n_1}}\;, \qquad
\hat{\a}=\hat{\a}_1 =\frac{1}{{\bar z}^{n_1}} \hat{\a}'\;, 
\label{A19}
\ee
where $\hat{\a}'$ is a $(0, 3)$-form independent of $z_1, {\bar z}_1$ and $d {\bar z}_1$.

Now we have all the ingredients to integrate by parts in~\eqref{A10}. Doing this once leads to
\be 
\l \sim \int_{{\mathbb C}^5} d^5 z \wedge \hat{\b}^{1} \wedge {\bar \pt} \Big(\frac{1}{p_2}\Big) + 
{\rm boundary} \   {\rm terms}\,. 
\label{A20}
\ee
We focus on the boundary terms in this expression for $|z_1|\to \infty$ and first note that 
\be 
\frac{\pt}{\pt {\bar z}_1}  \Big(\frac{1}{p_1}\Big) d {\bar z}_1 \wedge {\bar \pt} \Big(\frac{1}{p_2}\Big) = 
\frac{\pt}{\pt {\bar z}_1}  \Big(\frac{1}{p_1}\Big) d {\bar z}_1 \wedge {\bar \pt}_{\hat{1}} \Big(\frac{1}{p_2}\Big)\,, 
\label{A21}
\ee
where ${\bar \pt}_{\hat{1}} $ is the Dolbeault operator with the derivative over ${\bar z}_1$ omitted. Then the boundary term for $|z_1|\to \infty$ turns into
\be 
\int_{{\mathbb C}^4 \times \gamma_1} d^5 z \wedge \frac{\hat{\a}}{p_1} \wedge {\bar \pt}_{\hat{1}} \Big(\frac{1}{p_2}\Big)\Big|_{|z_1|\to \infty}\,. 
\label{A22}
\ee
In the limit of large $|z_1|$ we generically have $p_1 \sim z_1^{q_1^{1}} p_1^{\prime}$, $p_2 \sim z_1^{q_2^{1}} p_2^{\prime}$, 
where $p_1^{\prime}$, $p_2^{\prime}$ are holomorphic polynomials independent of $z_1$. Inserting this into Eq.~\eqref{A22} gives
\be 
\int_{{\mathbb C}^4 \times \gamma_1} d^5 z \wedge \frac{\hat{\a}}{z_1^{n_1+1}} \frac{1}{p_1^{\prime}}
\wedge {\bar \pt}_{\hat{1}} \Big(\frac{1}{p_2^{\prime}}\Big)\Big|_{|z_1|\to \infty}\,. 
\label{A23}
\ee
This integral is indeed zero because  $n_1 >0$ and $\hat{\a} \to 0$ at infinity. 
 
Finally, we need to perform the second integration by parts in the first term in Eq.~\eqref{A20}. As before, we focus on the boundary term for $|z_1|\to \infty$ which is given by
\be 
\int_{{\mathbb C}^4 \times \gamma_1} d^5 z \wedge \frac{\hat{\b}^{1}}{p_2} \Big|_{|z_1|\to \infty}
\sim 
\int_{{\mathbb C}^4 \times \gamma_1} d^5 z  \wedge 
\frac{d {\bar  z}_2 \wedge \dots \wedge  d {\bar  z}_5}{z_1^{n_1+1}  {\bar z}_1^{n_1} p_2^{\prime}} \Big|_{|z_1|\to \infty} =0\,. 
\label{A24}
\ee
%


\section{The coboundary map}\label{appB}

It is a well-known that for every short exact sequence of sheaves there is an associated long exact sequence in sheaf cohomology. A crucial ingredient in this correspondence is the co-boundary map whose construction can be found in standard textbooks, see for example~\cite{gh}, page 40. Since the co-boundary map plays an important role for our discussion in the main part of the paper, we now briefly review its construction.\\[2mm]
We start with the short exact sequence
\be 
0 \longrightarrow A \stackrel{g}{\longrightarrow} B  \stackrel{f}{\longrightarrow} C  \longrightarrow 0
\label{B1}
\ee
of sheaves $A$, $B$, $C$ and sheave morphisms $f$, $g$, satisfying $f\circ g=0$. The associated long exact sequence in cohomology has the form
\bea
 \cdots&\longrightarrow &  H^i (A) \stackrel{g}{\longrightarrow} H^i (B) \stackrel{f}{\longrightarrow} H^i (C)\nonumber \\
&\stackrel{\delta}{\longrightarrow} &H^{i+1} (A) \stackrel{g}{\longrightarrow} H^{i+1} (B) \stackrel{f}{\longrightarrow} H^{i+1} (C) 
\longrightarrow \dots\;.
\label{B2}
\eea
where $f$ and $g$ are the induced maps in cohomology and $\delta$ is the co-boundary map which needs to be constructed. To be in line with the main part of the paper, we will use the language appropriate for vector bundles, rather than more general sheaves, from now on.

To derive $\delta$, we start with a differential $(0, i)$-form $\nu \in H^i (C)$ taking values in $C$. Since the map $f: B \to C$ in~\eqref{B1} is surjective it follows that 
$\nu$ can always be written as  $\nu = f (\hat{\nu})$ for some form $\hat{\nu} \in \Omega^i (B)$. However, if $H^{i+1} (A) \neq 0$ the induced map $f: H^i (B) \to H^i (C)$ is not surjective which implies that the form  $\hat{\nu}$ is not necessarily closed.  Now we consider 
${\bar \pt} \hat{\nu} \in \Omega^{i+1} (B)$. We get 
\be 
f ({\bar \pt} \hat{\nu}) = {\bar \pt} (f ( \hat{\nu})) =  {\bar \pt} \nu =0\,, 
\label{B3}
\ee
where we have used the fact that the map $f$ is holomorphic. This implies that ${\bar \pt} \hat{\nu}$ is in the kernel of $f$
and by exactness of the sequence~\eqref{B2} it is in the image of $g$. That is, there exists an element $\hat{\o}\in 
\Omega^{i+1} (A)$ such that $g\hat{\o} = {\bar \pt} \hat{\nu}$. Moreover, since $g\bar\pt\hat{\o}=\bar\pt g\hat{\omega}=\bar\pt^2\hat{\nu}=0$ and $g$ is injective we have $\bar\pt\hat{\omega}=0$. Hence, $\hat{\omega}$ represents an element of $H^{i+1}(A)$ and we can define the co-boundary map by
\be
 \delta(\nu)=\hat{\omega}\; .
\ee 
In summary, the main features of the short exact sequence~\eqref{B1} and its long exact counterpart~\eqref{B2} that we will require are as follows. For a $(0,i)$--form $\nu\in H^i(C)$ and its image $\hat{\omega}=\delta(\nu)\in H^{i+1}(A)$ under the co-boundary map, there exist a $(0,i)$--form $\hat{\nu}\in \Omega^i(B)$ such that
\be 
\nu = f (\hat{\nu})\; ,\qquad {\bar \pt} \hat{\nu}=g \hat{\o}\; .
\label{B4}
\ee


\section{Harmonic line bundle-valued forms on ${\mathbb P}^n$}\label{appC}
One of the main ingredients of our calculation of Yukawa couplings is the explicit construction of bundle-valued forms, representing line bundle cohomologies on the ambient space. Since the ambient spaces under consideration are products of projective spaces it is sufficient to discuss a single projective space $\mathbb{P}^n$. For $\mathbb{P}^1$ this was done in Ref.~\cite{Blesneag:2015pvz}. Here we consider the generalisation to arbitrary projective spaces. 

We begin by setting up and reviewing standard facts about projective space including the Fubini-Study metric. One way to obtain a one-to-one correspondence between cohomology and forms is to focus on harmonic forms and we will do this relative to the Fubini-Study metric. Line bundles, their Chern connections and cohomology are the subject of the next two parts of the appendix. Most of this material can be found in standard text books, such as Refs.~\cite{gh,hartshorne,huybrechts}. Finally, we explain how harmonic line-bundle valued forms are related under multiplication. 

\subsection{Basics of projective space}
Complex projective space $\mathbb{P}^n$ is defined as the set of complex lines through the origin in $\mathbb{C}^{n+1}$. We denote coordinates on $\mathbb{C}^{n+1}$ by $x_{\alpha}$, where $\a=0,1,\ldots ,n$. The element of $\mathbb{P}^n$ given by the line through the origin and a point $(x_0,x_1,\cdots ,x_n)$ (with at least one $x_{\alpha}\neq 0$) is denoted by $(x_0:x_1:\cdots :x_n)\in \mathbb{P}^n$. The standard open patches on $\mathbb{P}^n$ are $U_{\alpha} = \lbrace (x_0:x_1:...:x_{n}) \,\vert\, x_{\alpha} \neq 0 \rbrace$, where $\a=0,\ldots ,n+1$, with associated charts $(U_{\alpha},\phi_{\alpha})$ and maps $\phi_{\alpha}: U_{\alpha} \rightarrow \mathbb{C}^n$ defined by $\phi_{\alpha}(x_0:x_1:...:x_{n})=(\x_0^\a,\x_1^\a,\ldots ,\widehat{\x_\a^\a},\ldots,\x_n^\a)$. Here, $\x_\m^\a=x_\m/x_\a$ are the coordinates on $\mathbb{C}^n$ and it is understood that $\x_\a^\a=1$ is discarded. For an overlap $U_{\alpha} \cap U_{\beta} \neq \emptyset$, the transition functions $\phi_{\beta \alpha } = \phi_{\beta}\circ \phi_{\alpha}^{-1}:\mathbb{C}^n \rightarrow \mathbb{C}^n $ takes the form $\xi_{\mu}^{\alpha} \mapsto \xi_{\mu}^{\beta} =\frac{x_{\alpha}}{x_{\beta}}\xi_{\mu}^{\alpha} $.

On each patch $U_\a$, the Fubini-Study K\"ahler potential can be written as
\be
\label{c1}
K_{\alpha} = \dfrac{i}{2 \pi} \textrm{ln}( \kappa_{\alpha})\;,\qquad \kappa_{\alpha} = \sum_{\mu = 0}^{n} \vert \xi^{\mu}_{\alpha}\vert^2\; .
\ee
The associated Fubini-Study K\"ahler form is given by
\be
 J = \partial \bar{\partial} K_{\alpha} 
\ee 
as usual and it is easy to check that this definition is independent of $\alpha$ on the overlaps and, hence, gives a globally defined form on $\mathbb{P}^n$. 
The above K\"ahler form is normalised such that 
\be
 \int_{\mathbb{P}^n}J^n=1\; .
\ee 
It will frequently be convenient to work on the patch $U_0=\mathbb{C}^n$ whose coordinates we also denoted by $z_\m=x_\m/x_0$, where $\m=1,\ldots ,n$ and we write $\k=\k_0=1+\sum_{\m=1}^n|z_\m|^2$.

\subsection{Line bundles on projective space}
The $k^{\rm th}$ power of the hyperplane bundle on $\mathbb{P}^n$ is denoted by $L=\mathcal{O}_{\mathbb{P}^n}(k)$. For each patch $U_\a$, a hermitian bundle metric on $L$ is given by 
\begin{eqnarray}
H_{\alpha}=\kappa_{\alpha}^{-k}\; . \label{Hdef}
\end{eqnarray}
On the patch $U_0$, we also write $H=H_0=\kappa^{-k}$. The associated Chern connection $\nabla^{0,1}=\bar{\partial}$ and $\nabla^{1,0}=\bar{\partial}+A_{\alpha}$ is specified by the gauge field
\be
A_{\alpha}= \partial \textrm{ln} \bar{H}_{\alpha} = - k \partial \textrm{ln} \kappa_{\alpha} = 2 \pi i k \partial K_{\alpha}
\ee
whose curvature $F_{\alpha}=d A_{\alpha} = - \partial \bar{\partial} \textrm{ln} \bar{H}_{\alpha}$ is explicitly given by
\be
 F_{\alpha} =  k \partial \bar{\partial} \textrm{ln} \kappa_{\alpha} = - 2 \pi i k \partial \bar{\partial} K_{\alpha} = - 2 \pi i k J\; .
\ee 
For the first Chern class of $L=\mathcal{O}_{\mathbb{P}^n}(k)$ this implies
\be
 c_1(\mathcal{O}_{\mathbb{P}^n}(k))=\dfrac{i}{2 \pi} F = kJ\; ,
\ee
as expected. 

\subsection{Line bundle cohomology}
The dimension of line bundle cohomology for a line bundle $\cK=\mathcal{O}_{\mathbb{P}^n}(k)$ is described by Bott's formula
\be
\label{bott}
h^q(\mathbb{P}^n,\mathcal{O}_{\mathbb{P}^n}(k))=\begin{cases} \dfrac{(n+k)!}{n!k!} & \textrm{ for } q=0, \textrm{ } k\geq 0 \\  \dfrac{(-k-1)!}{n! (-k-n-1)!} \textrm{ } & \textrm{ for } q=n, \textrm{ } k\leq -(n+1)\\ 0 & \textrm{ otherwise }  \end{cases}
\ee
This means that line bundles $\mathcal{O}_{\mathbb{P}^n}(k)$ in the ``gap" $-n+1<k<0$ only have trivial cohomologies while all other line bundles have precisely one non-trivial cohomology. For $k\geq 0$, this non-trivial cohomology is the zeroth cohomology with dimension given in the first row of Eq.~\eqref{bott}. For $k\leq (-n-1)$, on the other hand, only the highest, $n^{\rm th}$ cohomology is non-trivial with dimension given in the second row of Eq.~\eqref{bott}.

We would like to represent these cohomologies by line bundle valued $(0,q)$--forms which are harmonic relative to the Fubini-Study metric. Such forms $\nu_\a$ should, on each patch $U_\a$ satisfy the equations (see Ref.~\cite{Blesneag:2015pvz} for details)
\be
\label{2equations}
\bar{\partial} \nu_\a =0\; , \qquad \partial(\bar{H}_{\alpha} \ast \nu_\a)=0\; ,
\ee
where $H_\a$ is the hermitian bundle metric~\eqref{Hdef}. To solve these equations, we should distinguish the different cases displayed in the Bott formula~\eqref{bott}.
\begin{enumerate}
\item \underline{$\cK=\mathcal{O}_{\mathbb{P}^n}(k)$ with $k \geq 0$:}\\
 In this case $H^0(\mathbb{P}^n,\mathcal{O}_{\mathbb{P}^n}(k)$ is the only non-zero cohomology so we are looking for sections, that is harmonic $(0,0)$--forms. On the patch $U_0$ they are given by
 \be
 \nu_{(k)}=P_{(k)}(z_1, ..., z_n)\; ,
\ee
where $P_{(k)}$ are polynomials of degree $k$ in $z_\m$.  It is straightforward to check that these have the correct transition functions upon transformation to another patch. Note that the dimension of the space of degree $k$ polynomials in $n$ variables is indeed given by the first line in the Bott formula~\eqref{bott}, as required.
\item \underline{$\cK=\mathcal{O}_{\mathbb{P}^n}(k)$ with $-(n+1)<k<0$:}\\
In this case, all cohomologies vanish and there are no harmonic forms to construct.
\item \underline{$\cK=\mathcal{O}_{\mathbb{P}^n}(k)$ with $k\leq -(n+1)$:}\\
In this case,  $H^n(\mathbb{P}^n,\mathcal{O}_{\mathbb{P}^n}(k)$ is the only non-vanishing cohomology so we are looking for harmonic $(0,n)$--forms. It is straightforward to verify that, on the patch $U_0$, these can be written as
\be 
 \nu_{(k)} = \kappa^k P_{(k)}(\bar{z}_1,...,\bar{z}_n) d \bar{z}_1 \wedge ... \wedge d \bar{z}_n\; ,
\ee
where $P_{(k)}$ are polynomials of degree $-k-n-1$ in the $n$ variables $\bar{z}_\m$. Note that the dimension of this polynomial space equals the value in the second row of the Bott formula~\eqref{bott}, as it should.
\end{enumerate}
For uniformity of notation, in the following $P_{(k)}$ for $k\geq 0$ denotes a polynomial of degree $k$ in $z_\m$, while $P_{(k)}$ for $k\leq -n-1$ denotes a polynomials of degree $-k-n-1$ in $\bar{z}_\m$.

\subsection{Multiplication of harmonic forms}
Calculating Yukawa couplings requires performing wedge products of harmonic bundle-valued forms on $\mathbb{P}^n$ (or on products of projective spaces) and we would like to understand in detail how this works. For the case of $\mathbb{P}^1$ this has been dealt with in Ref.~\cite{Blesneag:2015pvz} and here we would like to discuss the generalisation to arbitrary projective spaces.

As we have seen, on $\mathbb{P}^n$, we have harmonic bundle-valued $(0,0)$-forms $\nu_{(k)}=P_{(k)}$ which represent the cohomology $H^0(\mathbb{P}^n,\cO_{\mathbb{P}^n}(k))$ for $k\geq 0$ and harmonic bundle-valued $(0,n)$ forms $\nu_{(k)}=\kappa^k P_{(k)} d \bar{z}_1 \wedge ... \wedge d \bar{z}_n$ which represent the cohomology $H^n(\mathbb{P}^n,\cO_{\mathbb{P}^n}(k))$ for $k\leq  -n-1$. Performing a wedge product between any two of those forms clearly produces a $\bar\pt$--closed form which is a representative of the appropriate cohomology. If this wedge product is between two harmonic $(0,0)$--forms the result is clearly again a harmonic $(0,0)$--form. However, the situation is more complicated for a product of a harmonics $(0,0)$--form and a harmonics $(0,n)$--form. The result is a $\bar\pt$--closed $(0,n)$--form which, however, is generally not harmonic. An obvious problem is to find the harmonic $(0,n)$--form in the same cohomology class as this product.

To discuss this in detail, we start with a harmonic $(0,0)$--form $p_{(\d)}$ representing a class in $H^0(\mathbb{P}^n,\mathcal{O}_{\mathbb{P}^n}(\delta))$ and a harmonic $(0,n)$--form
\be
 \nu_{(k-\delta)}=\kappa^{k-\delta} P_{(k-\delta)} d \overline{z}_1 \wedge ... \wedge d \overline{z}_n \label{nuindef}
\ee 
representing a class in $H^n(\mathbb{P}^n,\mathcal{O}_{\mathbb{P}^n}(k-\delta))$, where $k\leq -n-1$. The product $p_{(\d)}\nu_{(k-\d)}$ is $\bar\pt$--closed, but not generally harmonic, and defines a class in $H^n(\mathbb{P}^n,\mathcal{O}_{\mathbb{P}^n}(k))$ whose harmonic representative we denote by
\be
 \nu_{(k)}=\kappa^{k} Q_{(k)} d \overline{z}_1 \wedge ... \wedge d \overline{z}_n\; .\label{nuoutdef}
\ee 
This harmonic representative differs from the original product by an exact piece, so we have an equation of the form
\be 
\label{mainequation}
p_{(\delta)} \nu_{(k-\delta)} + \overline{\partial} s = \nu_{(k)}\; , 
\ee
where $s$ is a section of $\mathcal{O}_{\mathbb{P}^n}(k)$. It turns out, and will be shown below, that the correct Ansatz for $s$ is
\be
s = \kappa_0^{k-\delta+1}\left( S^{(1)} d \overline{z}_2 \wedge ... \wedge d \overline{z}_n - S^{(2)} d \overline{z}_1\wedge\overline{z}_3 \wedge ... \wedge d \overline{z}_n+...+ (-1)^{n-1} S^{(n)} d \overline{z}_1 \wedge  ... \wedge d \overline{z}_{n-1} \right) \; ,
\ee
where the  $S^{(i)}$ are multivariate polynomials of degree $\delta-1$ in $z_i$ and of degree $-k+\delta-n$ in $\bar{z}_i$. Eq.~\eqref{mainequation} can be solved by inserting the various differential forms including the most general polynomials of the appropriate degrees and then matching polynomials coefficients. In this way, given $p_{(\delta)} $ and $\nu_{(k-\delta)}$, both $s$ and $\nu_{(k)}$ can be determined as we will see below. While this is straightforward in principle the details are complicated. However, the main result can be stated in a simple way and we would like to do this upfront. It turns out that the polynomial $Q_{(k)}$ which determines $\nu_{(k)}$ is given by
\be
\tilde{Q}_{(k)} = c\, \tilde{p}_{(\delta)} \tilde{P}_{(k-\delta)} \quad\mbox{where}\quad c=\dfrac{(-k-1)!}{(-k+\delta -1 )!}\; . \label{mainres}
\ee
We recall that the tilde denotes the homogenous counterparts of the various polynomials, so all polynomials in the above equation depend on the homogeneous coordinates $x_\m$, where $\m=0,1,\ldots ,n$. The polynomial ``multiplication" on the RHS of this equation should be carried out by converting the coordinates $x_\m$ in $\tilde{p}_{(\delta)}$ into the partial derivatives $\pt/\pt \bar{x}_\m$ which then, in turn, act on $\tilde{P}_{(k-\delta)}$ which depends on $\bar{x}_\m$. Note that this leads to the correct degree required for the polynomial $\tilde{Q}_{(k)}$. This remarkably simple solution to Eq.~\eqref{mainequation} is the key to converting the calculation of Yukawa couplings into an ``algebraic" calculation. From this result, the wedge products of harmonic forms which appears in the Yukawa integral can simple be converted into polynomial multiplication, with the appropriate conversion of coordinates into partial derivatives, as discussed. Although $s$ is determined by Eq.~\eqref{mainequation},  we are unfortunately not aware of a formula for $s$ as simple as Eq.~\eqref{mainres}.\\[2mm]
In order to proof Eq.~\eqref{mainres}, we first note the derivative
\begin{eqnarray}
\overline{\partial} s & =& \kappa_0^{k-\delta+1} \left( \partial_{\overline{z}_1} S^{(1)}+\partial_{\overline{z}_2} S^{(2)}+...+ \partial_{\overline{z}_n} S^{(n)}\right) d \overline{z}_1 \wedge  ... \wedge d \overline{z}_{n}+\\
&& (k-\delta+1)\kappa_0^{k-\delta}\left(z_1 S^{(1)}+...+z_n S^{(n)}\right)d \overline{z}_1 \wedge  ... \wedge d \overline{z}_{n} .
\end{eqnarray}
Inserting this together with Eqs.~\eqref{nuindef} and \eqref{nuoutdef} into Eq.~\eqref{mainequation} leads to
\be
\label{maineq}
p P+ \kappa \sum^n_{i=1} \partial_{\overline{z}_i} S^{(i)} - (-k+\delta-1) \sum^n_{i=1} z_i S^{(i)} = \kappa^{\delta} Q\; .
\ee
Next, we should write out each of the polynomials explicitly. For each $S^{(i)}$ we have 
\be
\label{expansion1}
S^{(i)}= \sum_{\lbrace 0 \leq i_1+...+i_n \leq \delta -1 \rbrace} \sum_{ \lbrace 0 \leq j_1+...+j_n \leq - k+\delta-n \rbrace} c^{(i)}_{(i_1...i_n; j_1 ... j_n)}z_1^{i_1} ... z_n^{i_n} \overline{z}_1^{j_1}...\overline{z}^{j_n}_n 
\ee
with coefficients $c^{(i)}_{(i_1...i_n; j_1 ... j_n)}$ such that $(i_1,...,i_n; j_1, ... ,j_n)$ represents any index combination satisfying $0 \leq i_1+...+i_n \leq \delta -1$ and $0 \leq j_1+...+j_n \leq - k+\delta-n$. Similarly, we can expand the other polynomials as
\begin{eqnarray}
\label{expansion2}
p_{(\delta)}&=& \sum_{0\leq i_1+...+i_n \leq \delta} a_{i_1 ... i_n} z^{i_1}_1...z^{i_n}_n\\
\label{expansion3}
P_{(k-\delta)}&=&\sum_{0\leq j_1+...+j_n\leq -k+\delta-n-1} b_{j_1...j_n} \overline{z}_1^{j_1}...\overline{z}_n^{j_n} \\
\label{expansion4}
Q_{(k)}&=&\sum_{0\leq j_1+...+j_n\leq -k-n-1} q_{j_1...j_n} \overline{z}_1^{j_1}...\overline{z}_n^{j_n} .
\end{eqnarray}
A useful polynomials expansion of $\kappa = 1+z_1 \overline{z}_1+...+z_n \overline{z}_n$ is given by
\be
\label{expansion5}
\kappa^{\delta}=\sum_{0\leq i_1+...+i_n\leq \delta } \dfrac{\delta !}{i_1 ! i_2! ... i_n! (\delta-i_1-...-i_n)!} z_1^{i_1}...z_n^{i_n} \overline{z}_1^{i_1}...\overline{z}_n^{i_n}\; .
\ee
Now substituting the polynomials from Eq.~\eqref{expansion1}--\eqref{expansion5} into Eq.~\eqref{maineq}, one can derive the following identity, by extracting the coefficient of the $z^{i_1}...z^{i_1} \overline{z}_1^{i_1+j_1}...\overline{z}_n^{i_1+j_1}$ term
\begin{multline}
\label{bbv}
\dfrac{\delta!}{i_1!...i_n!(\delta-i_1-...-i_n)!}q_{j_1...j_n}=a_{i_1...i_n}b_{l_1...l_n}+\sum_{s=1}^{n} (l_s+1)c^{(s)}_{i_1...i_n;l_1...l_s+1...l_n}+\sum_{s=1}^n (k-\delta+1+l_s)c^{(s)}_{i_1...i_s-1,...,i_n;l_1...l_n}\\+ \sum_{s=1}^n \sum_{\substack{r=1 \\ r \neq s}}^n (l_s+1) c^{(s)}_{i_1...i_r-1...i_n;l_1...l_r-1...l_s+1...l_n}\; ,
\end{multline}
where we have denoted $l_s=i_s+j_s$, for all $s=1,...,n$. Note, however, that Eq.~\eqref{bbv} is true only if all $i_s$ are strictly positive and strictly smaller than $\delta-\sum^n_{r \neq s} i_r$. For $i_s = 0$ the $c^{(s)}_{i_1,...,i_s-1,...,i_n;l_1...l_n}$ term is not present, because the polynomial expansion of $S^{(s)}$ contains only positive exponents. For $i_s = \delta-\sum^n_{r \neq s} i_r$, the term $c^{(s)}_{i_1...i_n;l_1...,l_s+1,...,l_n}$ is missing, because it does not respect the summation rule. However, we can conventionally define all these unwanted $c^{(s)}$ coefficients to be zero, so that Eq.~\eqref{bbv} is valid for any $i_s \geq 0$.

In order to solve the above set of equations for $q_{j_1...j_n}$, it is useful to define the quantities
\be
\label{defofbeta}
\beta_{i_1 ... i_n} = \dfrac{(-k+\delta-n-1-(i_1+...+i_n)-(j_1+...+j_n))!}{(-k-n-1-(j_1+...+j_n))!}\dfrac{(i_1+j_1)!}{j_1!}...\dfrac{(i_n+j_n)!}{j_n!}\; ,
\ee
which satisfy the following combinatorial identity
\be
\label{tobeproven}
\sum_{0\leq i_1+...+i_n\leq \delta}\beta_{i_1 ... i_n}  \dfrac{\delta!}{i_1!...i_n! (\delta-i_1-...-i_n)!}=\dfrac{(-k+\delta -1 )!}{(-k-1)!}\; .
\ee
A proof of this identity can be found at the end of this appendix. Next, we multiply both sides of Eq.~\eqref{bbv} by $\beta_{i_1 ... i_n} $ and then sum over all indices $\{i_1,\ldots ,i_n\}$ with $0 \leq i_1+...+i_n \leq \delta$. This trick removes all coefficients $c^{(s)}$ from our equation, as a result of the identity
\begin{multline}
\label{relief}
\sum_{0 \leq i_1+...+i_n \leq \delta} \beta_{i_1 ... i_n} \biggl(\sum_{s=1}^{n} (l_s+1)c^{(s)}_{i_1...i_n;l_1...l_s+1...l_n}+\sum_{s=1}^n (k-\delta+1+l_s)c^{(s)}_{i_1...i_s-1,...,i_n;l_1...l_n}\\+ \sum_{s=1}^n \sum_{\substack{r=1 \\ r \neq s}}^n (l_s+1) c^{(s)}_{i_1...i_r-1...i_n;l_1...l_r-1...l_s+1...l_n}\biggr)=0\; .
\end{multline}
To see this, consider the weight $w$ of an arbitrary coefficient $c^{(n)}_{i_1...i_n; l_1...l_{s}+1...l_n}$ in the above sum, defined as
\be
w = (k-\delta +2+l_s) \beta_{i_1...i_s+1...i_n} + (l_s+1) \beta_{i_1...i_n} + (l_s+1)\sum_{\substack{r=1 \\ r \neq s}}^{n} \beta_{i_1...i_r+1...i_n}\; .
\ee
Starting from the definition of $\beta$ in Eq~\eqref{defofbeta} we notice that
\be
(l_s+1) \beta_{i_1...i_r+1...i_n} = (l_r+1)\beta_{i_1...i_s+1...i_n}\; , \quad \forall r \neq s\; .
\ee
Therefore the weight of $c^{(n)}_{i_1...i_n; l_1...l_{s}+1...l_n}$ becomes
\be
w=(k-\delta+\sum_r^n l_r + n+1) \beta_{i_1...i_s+1...i_n} + (l_s+1) \beta_{i_1...i_n} ,
\ee
which vanishes. Coming back to Eq.~\eqref{bbv}, we multiply with $\beta_{i_1 ... i_n}$ and sum over all $\lbrace i_1,...,i_n\rbrace$ with $0 \leq i_1+...+i_n \leq \delta$. This removes $c^{(i)}$ and leads to an equation for the coefficients of $Q$, namely
\be
\label{proofderiv}
 q_{j_1...j_n} = \dfrac{(-k-1)!}{(-k+\delta -1 )!} \sum_{0 \leq i_1+...+i_n \leq \delta } \beta_{i_1 ... i_n} a_{i_1...i_n}b_{l_1,...,l_n}\; .
\ee
We should now compare this result for $Q$, obtained by solving Eq.~\eqref{mainequation}, with the proposed solution~\eqref{mainres}. To this end, we convert all relevant polynomials into their homogeneous counterparts and also convert the coordinates in $\tilde{p}_{(\d)}$ into derivatives. This leads to
\begin{eqnarray}
\tilde{p}_{(\delta)}&=&\sum_{i_0+...+i_n = \delta} a_{i_1...i_n}\left( \dfrac{\partial}{\partial \overline{x}_0}\right)^{i_0}\left( \dfrac{\partial}{\partial \overline{x}_1}\right)^{i_1}...\left(\dfrac{\partial}{\partial \overline{x}_n}\right)^{i_n}\\
\tilde{P}_{(k-\delta)}&=&\sum_{j_0+ ...+j_n = -k+\delta-n-1} b_{j_1...j_n} \overline{x}_0^{j_0} \overline{x}_1^{j_1}... \overline{x}_n^{j_n} \\
\tilde{Q}_{(k)}&=&\sum_{ j_0 + ...+j_n= -k-n-1} q_{j_1...j_n} \overline{x}_0^{j_0} \overline{x}_1^{j_1}... \overline{x}_n^{j_n}\; .
\end{eqnarray}
Inserting this into the RHS of Eq.~\eqref{mainres} gives
\be
\tilde{p}_{(\delta)} \tilde{P}_{(k-\delta)} = \sum_{(i_0+...+i_n = \delta)} \sum_{(j_0 + ...+j_n= -k-n-1)} \underbrace{\dfrac{(i_0+j_0)!}{j_0!}...\dfrac{(i_n+j_n)!}{j_n!}}_{\beta_{i_1 ... i_n}} a_{i_1...i_n} b_{(i_1+j_1)...(i_n+j_n)} \overline{x}_0^{j_0} \overline{x}_1^{j_1}... \overline{x}_n^{j_n}
\ee
and inserting the result~\eqref{proofderiv} for the coefficients of $Q$ proofs Eq.~\eqref{mainres}.\\[2mm]
\underline{\textbf{Proof of of Eq.~\eqref{tobeproven}:}}
We start from the $n=1$ equation
\be
\label{n1case}
\sum_{i=0}^{\delta} \dfrac{(-k+\delta -2-i-j)!(i+j)!}{(-k-j-2)!j!} \dfrac{\delta !}{i! (\delta -i)!} = \dfrac{(-k+\delta -1)!}{(-k-1)!}\; ,
\ee
which can be proven by explicit calculation. It is then useful to write the sum over $n$ as
\be
\sum_{i_1=0}^{\delta} \sum_{i_2=0}^{\delta-i_1}...\sum_{i_n=0}^{\delta-i_1-...-i_{n-1}} \dfrac{(-k+\delta-n-1-\sum_s^n l_s)!}{(-k-n-1-\sum_s^n j_s)!} \dfrac{l_1 !}{j_1 !}... \dfrac{l_n !}{j_n !} \dfrac{\delta!}{i_1!...i_n!(\delta-i_1-...-i_n)!} ,
\ee
and to perform the summation step by step, starting from $i_n$ and ending with $i_1$, while using Eq.~\eqref{n1case} every time. For $i_n$, we use Eq.~\eqref{n1case} with $\delta_n=\delta - \sum_{s=1}^{n-1} i_s$ instead of $\delta$ and $k_n=k+n-1+\sum_{s=1}^{n-1} j_s$ instead of $k$ which leads to
\begin{multline}
\sum^{\delta-i_1-...-i_{(n-1)}}_{i_n=0} \dfrac{(-k+\delta-n-1-\sum_s^n l_s)!}{(-k-n-1-\sum_s^n j_s)!} \dfrac{l_n!}{j_n!} \dfrac{\delta !}{i_n! (\delta-i_1-...-i_n)!}=\\=\dfrac{(-k+\delta-n-\sum_{s=0}^{n-1} l_s)!}{(-k-n-\sum_{s=0}^{n-1} j_s)!} \dfrac{\delta!}{(\delta-\sum_{s=1}^{n-1}i_s)!}=\dfrac{(-k_{n-1}+\delta_{n-1}-2-l_{n-1})!}{(-k_{n-1}-2-j_{n-1})!}\dfrac{\delta!}{(\delta_{n-1}-i_{n-1})!}
\end{multline}
After performing all the sums, we obtain the required result~\eqref{tobeproven}.


\end{document}